# The Challenge of Spin-Orbit-Tuned Ground States in Iridates: A Key Issues Review

*Gang Cao*

*Department of Physics, University of Colorado at Boulder, Boulder, CO 80309, USA*

*Pedro Schlottmann*

*Department of Physics, Florida State University, Tallahassee, FL 32306, USA*

## Abstract

Effects of spin-orbit interactions in condensed matter are an important and rapidly evolving topic. Strong competition between spin-orbit, on-site Coulomb and crystalline electric field interactions in iridates drives exotic quantum states that are unique to this group of materials. In particular, the "$J_{eff} = \frac{1}{2}$" Mott state served as an early signal that the combined effect of strong spin-orbit and Coulomb interactions in iridates has unique, intriguing consequences. In this *Key Issues Review*, we survey some current experimental studies of iridates. In essence, these materials tend to defy conventional wisdom: absence of conventional correlations between magnetic and insulating states, avoidance of metallization at high pressures, "S-shaped" I-V characteristic, emergence of an odd-parity hidden order, etc. It is particularly intriguing that there exist conspicuous discrepancies between current experimental results and theoretical proposals that address superconducting, topological and quantum spin liquid phases. This class of materials, in which the lattice degrees of freedom play a critical role seldom seen in other materials, evidently presents some profound intellectual challenges that call for more investigations both experimentally and theoretically. Physical properties unique to these materials may help unlock a world of possibilities for functional materials and devices. We emphasize that, given the rapidly developing nature of this field, this *Key Issues Review* is by no means an exhaustive report of the current state of experimental studies of iridates.

# Table of Contents

## I. Introduction

It is now apparent that novel materials, which often exhibit surprising or even revolutionary physical properties, are necessary for critical advances in technologies that affect the everyday lives of people. For example, a number of discoveries involving superconducting and magnetic materials, polymers and thin-film processing have underpinned the development of novel medical diagnostic tools, personal electronic devices, advanced computers, and powerful motors and actuators for automobiles. Transition metal oxides have recently been the subject of enormous activity within both the applied and basic science communities. However, the overwhelming balance of interest was devoted to *3d*-elements and their compounds for many decades. Scientists, confronted with ever-increasing pace and competition in research, are beginning to examine the remaining "unknown territories" located in the lower rows of the periodic table of the elements. Although the rare earth and light actinide elements have been aggressively studied for many decades, the heavier *4d*- and *5d*-elements and their oxides have largely been ignored until recently. The reduced abundance and increased production costs for many of these elements have certainly discouraged basic and applied research into their properties. What has not been widely appreciated, however, is that *4d*- and *5d*-elements and their compounds exhibit unique competitions between fundamental interactions that result in physical behaviors and empirical trends that markedly differ from their *3d* counterparts.

Recently, iridium oxides or iridates have attracted growing attention, due to the influence of strong spin-orbit interactions (SOI) on their physical properties; these effects were largely ignored in theoretical treatments of *3d* and other materials of interest (except for the rare earth and actinide classes). Traditional arguments suggest that iridates should be more metallic and less magnetic than materials based upon 3*d* and 4*f* elements, because *5d*-electron orbitals are more extended in space, which increases their electronic bandwidth (**Fig.1a**). This conventional wisdom conflicts with early observations of two empirical trends inherent in iridates such as the Ruddlesden-Popper phases, $Sr_{n+1}Ir_nO_{3n+1}$ (n = 1 and 2; n defines the number of Ir-O layers in a unit cell) [1,2,3,4,5] and hexagonal BaIrO₃ [6]. First, complex magnetic states occur with high critical temperatures but unusually low ordered moments. Second, "exotic" insulating states, rather than metallic states, are commonly observed [1,2,3,4,5,6] (see **Table 1**).

The early observations during the 1990's and early 2000's [1,2,3,4,5,6], for example, signaled physics unique to the *5d*-based materials and motivated extensive investigations in recent years, which eventually led to the recognition that a rare interplay of on-site Coulomb repulsion, U, crystalline fields and strong SOI has unique, intriguing consequences in iridates. The most profound manifestation of such an interplay is characterized by the $J_{eff}=1/2$ Mott state in $Sr_2IrO_4$, which was first identified in 2008 [7,8,9]. This quantum state represents novel physics and has since generated a surge of interest in this class of materials.

**Table 1.** *Examples of Layered Iridates*

| System | Néel Temperature (K) | Ground State |
|---|---|---|
| $Sr_2IrO_4$ (n =1) | 240 | Canted AFM insulator |
| $Sr_3Ir_2O_7$ (n =2) | 285 | AFM insulator |
| $BaIrO_3$ | 183 | Canted AFM insulator |

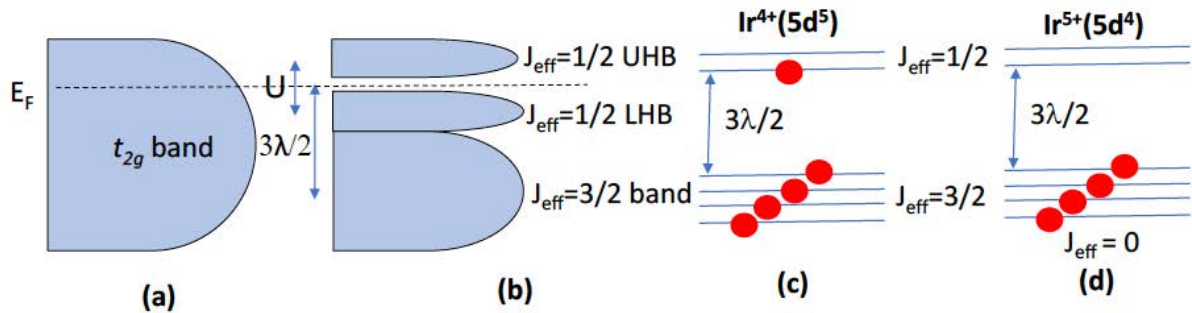

***Fig.1.*** *Band schematic: **(a)** The traditionally anticipated broad $t_{2g}$ band for 5d-electrons; **(b)** The splitting of the $t_{2g}$ band into $J_{eff}=1/2$ and $J_{eff}=3/2$ bands due to SOI; **(c)** $Ir^{4+}(5d^5)$ ions provide five 5d-electrons, four of them fill the lower $J_{eff} = 3/2$ bands, and one electron partially fills the $J_{eff} = 1/2$ band where the Fermi level $E_F$ resides; and **(d)** For $Ir^{5+}(5d^4)$ ions, four 5d-electrons fill the $J_{eff}=3/2$ bands, leading to a singlet $J_{eff} = 0$ state for the strong SOI limit.*

It is now recognized that the strong SOI along with U can drive novel narrow-gap Mott states in iridates. ***The SOI is a relativistic effect that is proportional to $Z^2$*** (Z is the atomic number; e.g., Z= 29 and 77 for Cu and Ir, respectively) [10, 11]. It needs to be pointed out that for the outer electrons, the strength of SOI scales with $Z^2$, rather than $Z^4$ although the $Z^4$–dependence is more often cited in the literature [11]. This is because the screening of the nuclear charge by the core electrons renders the effective $Z^2$-dependence of SOI for the large Z outer electrons; the $Z^4$-dependence of SOI is for the unscreened hydrogenic wavefunctions. The SOI is approximately 0.4 eV in iridates (compared to ~ 20 meV in *3d* materials), and splits the $t_{2g}$ bands into states with $J_{eff} = 1/2$ and $J_{eff} = 3/2$, the latter having lower energy [7] (**Fig. 1b and Table 2**). Since $Ir^{4+}$ ($5d^5$) ions provide five 5d-electrons to bonding states, four of them fill the lower $J_{eff} = 3/2$ bands, and one electron partially fills the $J_{eff} = 1/2$ band, where the Fermi level $E_F$ resides (**Fig.1c**). A combined effect of the strong SOI and U opens an energy gap $\Delta$ of the Mott type in the Jeff=1/2 band supporting the insulating state in the iridates (**Fig.1b**) [7]. The splitting between the $J_{eff} = 1/2$ and $J_{eff} = 3/2$ bands narrows as the dimensionality (i.e., n) increases in $Sr_{n+1}Ir_nO_{3n+1}$, and the two bands progressively broaden and contribute to the density of states (DOS) near the Fermi surface. In particular, the bandwidth, W, of the $J_{eff} = 1/2$ band increases from 0.48 eV for n = 1 to 0.56 eV for n = 2 and 1.01 eV for n = ∞ [8]. The ground state evolves with decreasing charge gap $\Delta$, from a robust insulating state for $Sr_2IrO_4$ (n = 1) to a metallic state for $SrIrO_3$ (n = ∞) as n increases. While the electron hopping occurs via direct d-d overlap and/or through the intermediate oxygen atoms between Ir ions, it is emphasized that the lattice plays a critical role that is seldom seen in other materials in determining the ground state, as discussed below.

The SOI-driven $J_{eff}$=1/2 model captures the essence of physics of many iridates with tetravalent $Ir^{4+}$($5d^5$) ions (see **Fig.1c**). The $J_{eff}$=3/2 states can accommodate four 5d electrons, leaving the fifth electron and one hole in the $J_{eff}$=1/2 band because there are six states (**Fig. 1c**). This can lead to the opening of a Mott gap. For $Ir^{5+}$ ($5d^4$) ions, on the other hand, the $J_{eff}$=3/2 band exhausts the available electrons and the $J_{eff}$=1/2 states remain empty (see **Fig. 1d**). However, it is a single-particle approach that is limited to situations where Hund's rule interactions among the electrons can be neglected and it may break down when non-cubic crystal fields, which are not taken into account in this model, become comparable to the SOI. This may result in a mixing of $J_{eff} = 1/2$ and $J_{eff} = 3/2$ states, altering isotropic wavefunctions that the model is based upon. Indeed, the Hund's rule interactions are more critical for irdates with $Ir^{5+}$ ($5d^4$) ions, and can mix $J_{eff} = 1/2$ and $J_{eff} = 3/2$ states, resulting in a qualitatively different picture (more discussion in **Section IV**). One such breakdown occurs in $Sr_3CuIrO_6$ due to strong non-octahedral crystal fields [12].

Nevertheless, a wide array of novel phenomena in iridates have been revealed in recent years [e.g. [13,14,15,16,17,18,19,20,21,22,23,24,25], and references therein]. It has become apparent that materials with such a delicate interplay between SOI, U and other competing interactions offer wide-ranging opportunities for the discovery of new physics and development of new devices, and it is not surprising that the physics of iridates is one of the most important topics in contemporary condensed matter physics. A great deal of recent theoretical work has appeared in response to early experiments on iridates, such as spin liquids in hyper-kagome structures [20,21], superconductivity [22,23,24,25,26,27,28], Weyl semimetals with Fermi arcs, axion insulators [29], topological insulators, correlated topological insulators [17], Kitaev modes, 3D spin liquids with Fermionic spinons, topological semimetals [13,17,30,31,32,33,34,35,36,37,38], etc. These and many other theoretical studies have advanced our understanding of *5d*-based materials and motivated enormous activity in search of novel states in iridates. It is particularly intriguing that many proposals have met limited experimental confirmation thus far, which makes iridates even more unique and challenging. We note that the SOI is a strong competitor with U and other interactions, which creates an entirely new hierarchy of energy scales (**Table 2**); the lack of experimental confirmation underscores a critical role of subtle structural distortions that may dictate the low-energy Hamiltonian [13] and need to be more thoroughly addressed both experimentally and theoretically.

**Table 2.** *Comparison between 3d and 4d/5d Electrons*

| Electron Type | U(eV) | λ(eV) | Spin State | Interactions | Phenomena |
|:---:|:---:|:---:|:---:|:---:|:---:|
| *3d* | 5-7 | 0.01-0.1 | *High* | $U > CF > \lambda$ | *Magnetism/HTSC* |
| *4d* | 0.5-3 | 0.1-0.3 | *Low/Intermediate* | $U \sim CF > \lambda$ | *p-wave SC* |
| *5d* | 0.4-2 | 0.1-1 | *Low* | $U \sim CF \sim \lambda$ | $J_{eff}$=1/2 Mott State |

*Note: SC=superconductivity; Hund's rule coupling is significant for 4d- and 5d-electrons.*

In this *Review*, we will discuss some underlying properties of certain representative iridates, namely, layered perovskite or Ruddlesden-Popper iridates (**Section II**), honeycomb and other geometrically frustrated lattices, all of which feature $Ir^{4+}$($5d^5$) ions (**Fig.1c**) (**Section III**), and double perovskite iridates with $Ir^{5+}$($5d^4$) ions (see **Fig.1d**) (**Section IV**). We review the current status of experimental studies of these materials and challenges that this class of materials presents. We conclude the *Review* with a list of outstanding issues and our outlook of the field, which hopefully could help stimulate more discussions (**Section V**). It is emphasized that this *Review* is intended to offer only a brief (thus incomplete) survey of some exemplary *experimental observations* and challenges, and it is by no means an exhaustive one. Given the nature of the broad and rapidly evolving field, it is inevitable that we may miss some important results in this field. In addition, this *Review* primarily focuses on properties of bulk single crystals, as a result, a large number of excellent studies of thin films and heterostructures of iridates, for example Refs **[39,40,41,42,43,44,45,46]**, are not included in this article. Since lattice properties are so critical to ground states of iridates, epitaxial thin films with varied strain and/or heterostructures offer a unique, powerful tool for tuning ground states of iridates. This is evidenced in Refs. [39,40,41,42,43,44,45,46] and many other recent publications. There are also excellent review articles emphasizing theoretical aspects of iridates [17,18].

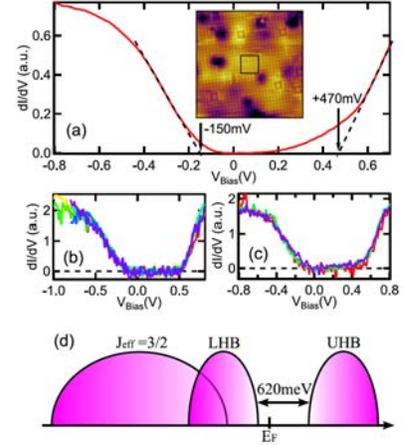

**Fig. 2.** *$Sr_2IrO_4$: The 620 meV energy gap.* **(a)** *The local density of states (LDOS) measured over the $2 \times 2$ $nm^2$ area indicated by the square in the image (inset). The dashed lines are drawn to indicate the band edges at $-150$ mV and $+470$ mV.* **(b)** *and* **(c)** *LDOS taken at locations away from defects, (b) without and (c) with the gradual increase.* **(d)** *Diagram showing energy bands with two important features: the 620 meV insulating gap and the overlap between the lower Hubbard band (LHB) and the $J_{eff} = 3/2$ band (Ref.87).*

**Table 3.** *Some Existing Iridates and Exemplary Phenomena*

| Structure | Compound | Ir Ion | Exemplary Phenomena |
|---|---|---|---|
| *Layered Perovskite* | *1. $Sr_2IrO_4$\* 2. $Sr_3Ir_2O_7$\* 3. $SrIrO_3$ 4. $Ba_2IrO_4$ \* Dopants* | $Ir^{4+}(5d^5)$ | *1. $J_{eff}$= 1/2 insulator; $T_N$=240 K [1,2,3,4], S-shaped IV curves [4] \** *2. $J_{eff}$ = 1/2 or band insulator; $T_N$=285 K [5], confined metal at 60 GPa [47] \** *3. Paramagnetic (semi-)metal (high pressure phase) [8,48]* *4. $J_{eff}$= 1/2 insulator; $T_N$~240 K (high pressure phase) [49]* *\* Dopants: K [50], Ca [51], Mn [52], Ru [53,54], Rh [55], La [1], Eu [56], Tb [57]* |
| *Hexagonal Perovskite* | *1. $BaIrO_3$, 2. $SrIrO_3$ 3. $Ca_5Ir_3O_{12}$ 4. $Ca_4IrO_6$ 5. $Ba_3IrTi_2O_9$ 6. $Ba_3NdIr_2O_9$ 7. $Ba_3LiIr_2O_8$ 8. $Ba_3NaIr_2O_8$* | $Ir^{4+}(5d^5)$  $Ir^{4.5+}$ $Ir^{4.5+}$ $Ir^{4.5+}$ | *1. $J_{eff}$ = 1/2 Mott insulator; $T_N$=183 K; CDW, S-shaped IV curves [6]* *2. Nearly ferromagnetic metal [58]* *3. Insulator; $T_N$=12 K [59]* *4. Insulator; $T_N$=6 K [59]* *5. Insulator; no long-range order [60]* *6. Magnetic insulator; $T_N$=20 K [61]* *7. AFM order at $T_N$ = 75 K [62]* *8. AFM order at $T_N$ = 50 K [62]* |
| *Honeycomb* | *1. $Na_2IrO_3$ 2. $\alpha$-$Li_2IrO_3$ 3. $\beta$-$Li_2IrO_3$ 4. $\gamma$-$Li_2IrO_3$ 5. $Na_4Ir_3O_8$* | $Ir^{4+}(5d^5)$ | *1. Zig-Zag Magnetic order; $T_N$=18 K [32,33,32,33]* *2. Incommensurate order [63]* *3. Incommensurate order [64,65]* *4. Incommensurate order [66,67]* *5. Spin liquid state [20]* |
| *Pyrochlore* | *1. $Bi_2Ir_2O_7$ 2. $Pb_2Ir_2O_7$ 3. $RE_2Ir_2O_7$ RE=Rare Earth Ion* | $Ir^{4+}(5d^5)$ | *1 and 2: Metallic states; Strong magnetic instability, etc. [68,69,70].* *3.Metal-insulator transition, insulating states, spin liquid, RE ionic size dependence [71,72]* |
| *Double Perovskite* | *1. $Sr_2YIrO_6$ 2. $Ba_2YIrO_6$ 3. $Sr_2REIrO_6$ 4. $Sr_2CoIrO_6$ 5. $Sr_2FeIrO_6$ 6. $La_2ZnIrO_6$ 7. $La_2MgIrO_6$ 8. $RE_2MIrO_6$ M=Mg, Ni* | $Ir^{5+}(5d^4)$  $Ir^{4+}(5d^5)$ | *1. Weak AFM state, correlated insulator [73]* *2. Insulator [74,75], AFM insulator [197]* *3. Magnetic insulator [73]* *4. Magnetic metal, $T_{N1}$=60 K, $T_{N2}$=120 K [76]* *5. Magnetic insulator, $T_N$=60 K [77]* *6. Weak ferromagnetic insulator, $T_N$=7.5 K [78]* *7. Weak ferromagnetic insulator, $T_N$=12 K [78]* *8. Weak AFM ground state with canted spin [79]* |
| *Post-Perovskite* | *$NaIrO_3$ $CaIrO_3$* | $Ir^{5+}(5d^4)$ $Ir^{4+}(5d^5)$ | *Paramagnetic insulator [80]* *Quasi-one-dimensional antiferromagnet [81,82]* |
| *Others* | *$Ba_5AlIr_2O_{11}$* | $Ir^{4.5+}$ | *1. Spin-½/dimer chain, charge and magnetic orders [83]* |

| | $Sr_3NiIrO_6$ | $Ir^{4+}(5d^5)$ | 2.One-dimensional chains and large coercivity [84] |
|---|---|---|---|

We tabulate a number of existing iridates in **Table 3**. These iridates are grouped according to their underlying crystal structures. Exemplary phenomena along with valence states of the iridates are also listed. There are a few general remarks on these iridates: With a few exceptions (e.g., $SrIrO_3$, $Bi_2Ir_2O_7$, and $Pr_2Ir_2O_7$), most iridates feature insulating and magnetic ground states; while the insulating states of these materials may vary in detail, they are due largely to the combined effect of SOI and U. Magnetic moments are in general weak, often a fraction of one Bohr magneton per Ir ion (The local moment for $J_{eff}=1/2$ state has 1 $\mu_B$ but covalency of Ir-O bonding reduces the moment). Furthermore, no first-order transitions are discerned thus far, and insulator-metal transitions tend to be gradual and continuous in this class of materials. A rotation of the $IrO_6$-octahedra about the c-axis is commonplace but tilting of the $IrO_6$-octahedra is rare in layered perovskite iridates.

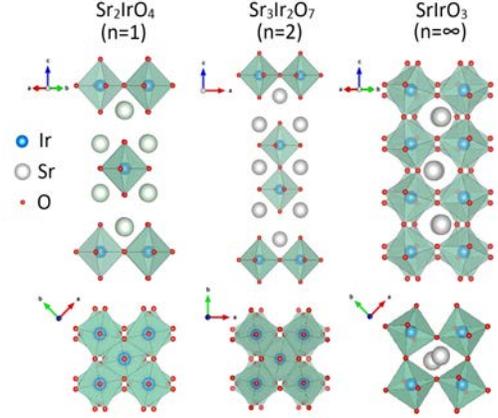

*Fig. 3. The crystal structure of layered perovskites $Sr_{n+1}Ir_nO_{3n+1}$ for n=1, 2 and ∞. Note that the lower three panels illustrate the rotation of $IrO_6$ octahedra about the c-axis for $Sr_2IrO_4$, $Sr_3Ir_2O_7$, and $SrIrO_3$, respectively.*

## II. Survey of Physical Properties of Ruddlesden-Popper Iridates, $Sr_{n+1}Ir_nO_{3n+1}$ (n=1, 2, and ∞)

### II.1. $Sr_2IrO_4$ (n=1)

$Sr_2IrO_4$ is the archetype $J_{eff}=1/2$ Mott insulator with a Néel temperature $T_N = 240$ K [1,2,3,4,7,9], an energy gap $\Delta \leq 0.62$ eV (see **Fig. 2**) [85,86,87] and a relatively small magnetic coupling energy of 60-100 meV [24,86]. It is perhaps the most extensively studied iridate thus far. The distinct energy hierarchy featuring a strong SOI and its structural, electronic and magnetic similarities to those of the celebrated $La_2CuO_4$ (i.e., $K_2NiF_4$ type, one hole per Ir or Cu ion, pseudospin- or spin-1/2 AFM, etc.) have motivated a large number of experimental and theoretical investigations on $Sr_2IrO_4$ in recent years.

## A. Critical Structural Features

A unique and important structural feature of $Sr_2IrO_4$, which has critical implications for the ground state, is a rotation of the $IrO_6$-octahedra about the c-axis by ~11°, which results in a larger unit cell volume by a factor $\sqrt{2}$ x $\sqrt{2}$ x 2. It is commonly thought that $Sr_2IrO_4$ crystallizes in a tetragonal structure with space-group $I4_1/acd$ (No. 142) with $a = b$ = 5.4846 Å and $c$ = 25.804 Å at 13 K [1,2,3], as shown in **Fig. 3**. More recently, studies of neutron diffraction and second-harmonic generation (SHG) of single-crystal $Sr_2IrO_4$[88,89] reveal structural distortions and forbidden reflections such as (1, 0, 2n+1) for the space group $I4_1/acd$ over a wide temperature interval, 4 K < T < 600 K. These results indicate the absence of the c- and d-glide planes, leading to a further reduced structural symmetry with a space group $I4_1/a$ (No. 88). One defining characteristic of $Sr_2IrO_4$ and other iridates is that the strong SOI strongly couples physical properties to the lattice degrees of freedom [13,50,88,90], which is seldom seen in other materials [91,92].

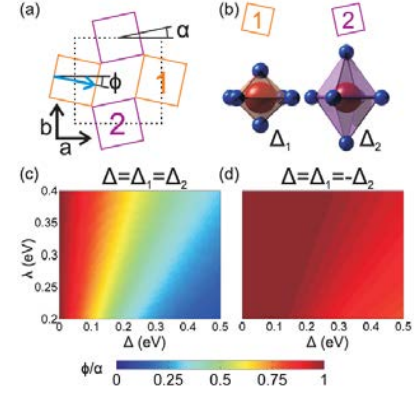

**Fig. 4. $Sr_2IrO_4$:** *(a) Illustration of an $IrO_2$ plane. The oxygen octahedra rotate about the c-axis by a creating a two-sublattice structure. The magnetic moments couple to the lattice and exhibit canting angles φ. (b) An unequal tetragonal distortion ($\Delta_1$ and $\Delta_2$) on the two sublattices as required by the $I4_1/a$ space group. (c) The ratio φ/α as a function of both SOI λ and Δ calculated for the case of uniform and (d) staggered ($\Delta_1 = -\Delta_2$) tetragonal distortion assuming $U$ = 2.4 eV, Hund's coupling $J_H$ = 0.3 eV, hopping $t$ = 0.13 eV, and α = 11.5° (Ref.94).*

The relationship between the rotation of the $IrO_6$ octahedra and magnetic moment canting in the iridates was first discussed in **Ref. 13** in which a theoretical model proposed a strong magnetoelastic coupling in $Sr_2IrO_4$, and a close association between the magnetic moment canting and the ratio of the lattice parameter of the c-axis to the a-axis [13], as a consequence of the strong SOI. Indeed, the strong locking of the moment canting to the $IrO_6$ -rotation (by 11.8(1)°) is experimentally manifest in studies of x-ray resonant scattering [93] and SHG [94]. In particular, the SHG study indicates that the *I41/a* space group requires a staggering of the sign of the tetragonal distortion ($\Delta_1$ and $\Delta_2$), which helps explain the magnetoelastic locking, as illustrated in **Fig. 4** [94].

As discussed below, the rotation of IrO$_6$ octahedra, which corresponds to a distorted in-plane Ir1-O2-Ir1 bond angle, plays an extremely important role in determining the electronic and magnetic structures. The Ir1-O2-Ir1 bond angle can be tuned via magnetic field [50], high pressure [95], electric field [90] and epitaxial strain [39]. The lattice properties not only make the ground state readily tunable but also provide a new paradigm for development of functional materials and devices.

Remarkably, that the 13(1)° canting of the moments away from the *a*-axis closely tracks the staggered rotation of the IrO$_6$ octahedra[93,94] sharply contrasts the behavior of *3d* oxides [96], in which structural distortions are noticeably decoupled with magnetic moment canting.

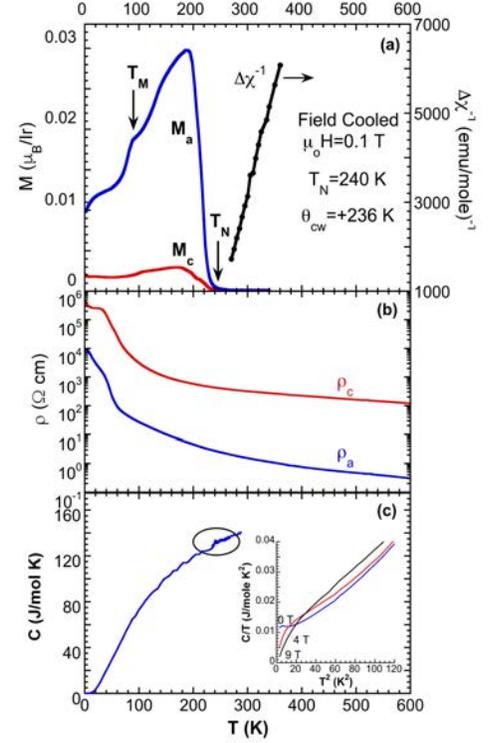

**Fig.5.** *Sr$_2$IrO$_4$: (a) The temperature dependence of magnetization for the a- and c-axis, $M_a$ and $M_c$, at 0.1 T and the inverse magnetic susceptibility $\Delta\chi^{-1}$(the right scale), (b) the electrical resistivity for the a- and c-axis, $\rho_a$ and $\rho_c$, and (c) the specific heat, C. The inset: C/T vs. $T^2$ at a few magnetic fields. $T_M$ marks a magnetic anomaly near 100 K (Ref. 50).*

## B. *Magnetic Properties*

Early experimental studies suggested that Sr$_2$IrO$_4$ was a weak ferromagnet with a Curie temperature at 240 K primarily because the temperature dependence of magnetization, along with a positive Curie-Weiss temperature, $\theta_{CW}$ = +236 K, appeared to be consistent with that of a weak ferromagnet (see **Fig.5a**) [2-5]. It is now understood that the observed weak ferromagnetic behavior arises from an underlying canted antiferromagnetic (AFM) order [88]. This also raises a question as to why $\theta_{CW}$ (= +236 K), which is extrapolated from the inverse susceptibility $\Delta\chi^{-1}$ ($\Delta\chi = \chi(T) - \chi_o$, where $\chi_o$ is a T-independent contribution), is positive in the presence of the AFM ground state (**Fig.5a**) [4,50,97]. Results of more recent x-ray scattering and neutron diffraction investigations of single-crystal $Sr_2IrO_4$ confirm that the system indeed undergoes a long-range AFM transition at 224(2) K with an ordered moment of 0.208(3) $\mu_B$/Ir and a canted magnetic configuration within the basal plane [88]. The magnetic configuration illustrated in **Figs. 6b-d** shows that magnetic moments are projected along the *b*-axis with a staggered ↓ ↑ ↑ ↓ pattern along the *c*-axis. Remarkably, these moments deviate 13(1)° away from the *a*-axis **(Fig. 6d)**, indicating that the magnetic moment canting rigidly tracks the staggered rotation of the $IrO_6$ octahedra discussed above. This strong coupling is a signature feature of the iridates in general and sharply contrasts to the situation in *3d* oxides, such as $Ca_3Mn_2O_7$, where a collinear magnetic structure exists in a strongly distorted crystal structure [96]. Nevertheless, the magnetic canting results in 0.202(3) and 0.049(2) $\mu_B$/Ir-site for the *a*-axis and the *b*-axis, respectively **(Fig.6)** [88]. For comparison, the ordered moment extrapolated from the magnetization, which measures net magnetic moment, is less than 0.08 $\mu_B$/Ir within the basal plane [97].

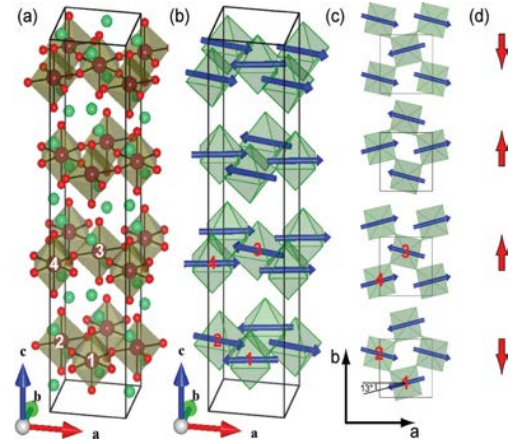

*Fig.6. $Sr_2IrO_4$: (a) The crystal structure of $Sr_2IrO_4$. Each $IrO_6$ octahedron rotates 11.8° about the c-axis. The Ir atoms of the non-primitive basis are labeled 1, 2, 3, and 4 plus the body centering translation (1/2,1/2,1/2). (b) The refined magnetic structure from single-crystal neutron diffraction measurements. (c) The same magnetic moment configuration projected on the basal planes. (d) The net moment projected along the b- axis for individual layers (Ref. 88).*

Indeed, a muon-spin rotation (μSR) study reports a low frequency mode that corresponds to the precession of weak ferromagnetic moments arising from a spin canting and a high frequency mode resulting from the precession of the antiferromagnetic sublattices [98]. Another study indicates a small energy gap for the AFM excitations, 0.83 meV, suggesting an isotropic Heisenberg dynamics [99]. Remarkably, a high-resolution inelastic light (Raman) scattering study of the low-energy magnetic excitation spectrum of $Sr_2IrO_4$ shows that the high-field (>1.5 T) in-plane spin dynamics is isotropic and governed by the interplay between the applied field and the small in-plane ferromagnetic spin components induced by the Dzyaloshinskii-Moriya interaction. However, the spin dynamics of $Sr_2IrO_4$ at lower fields (H <1.5 T) exhibits important effects associated with interlayer coupling and in-plane anisotropy, including a spin-flop transition in $Sr_2IrO_4$ that occurs either discontinuously or via a continuous rotation of the spins, depending on the in-plane orientation of the applied field. These results show that in-plane anisotropy and interlayer coupling effects play important roles in the low-field magnetic and dynamical properties of $Sr_2IrO_4$ [56]. It is also found that $Sr_2IrO_4$ (as well as $Sr_3Ir_2O_7$) exhibits pronounced two-magnon Raman scattering features and that the SOI might not be strong enough to quench the orbital dynamics in the paramagnetic state [100].

A close examination of the low-field magnetization, M(T), reveals two additional anomalies at $T_M \approx$ 100 K and 25 K (not marked) in $M_a(T)$ and $M_c(T)$ (see **Fig. 5a**), suggesting a moment reorientation below 100 K. The existence of the reorientation is further corroborated by the μSR study of $Sr_2IrO_4$ in which two structurally equivalent muon sites experience distinct local magnetic fields below 100 K characterized by the development of a second precession signal that is fully established below 20 K [98]. This behavior, which is likely a result of the balance of competing energies that changes with temperature, correlates with a change in the Ir1-O2-Ir1 bond angle that leads to the gradual reorientation near 100 K. This reorientation of the moments may be at the root of the unusual magnetoresistivity [50] and giant magnetoelectric behavior [97] that does not depend on the magnitude and spatial dependence of the magnetization, as conventionally anticipated.

## C. *Transport Properties*

The electrical resistivity of $Sr_2IrO_4$ for the *a*- and *c*-axis, $\rho_a$ and $\rho_c$, exhibits insulating behavior throughout the entire temperature range measured up to 600 K, as shown in **Fig. 5b**. The anisotropy, $\rho_c/\rho_a$, is significant, ranging from $10^2$ to $10^3$ although it is much smaller than $10^4 - 10^5$ for La$_2$CuO$_4$ because of the extended nature of the *5d*-electrons.

Both $\rho_a$ and $\rho_c$ exhibit anomalies at $T_M$ ($\approx$100 K) and 25 K, but conspicuously not at $T_N$ (=240 K), as shown in **Figs. 5a** and **5b**. It has now become clear that this is a signature behavior of Sr$_2$IrO$_4$ in which transport properties exhibit no discernable anomaly corresponding to the AFM transition at $T_N$=240 K [4,50,97]. In addition, the observed specific heat anomaly $|\Delta C|$ is tiny, ~ 4 mJ/mole K for Sr$_2$IrO$_4$ (**Fig.5c**), in spite of its robust, long-range magnetic order at $T_N$= 240 K. It is worth mentioning that C(T) below 10 K is predominantly proportional to T$^3$ at $\mu_o$H = 0 and 9 T (**Fig. 5c** inset), due to a Debye-phonon and/or magnon contributions from the AFM ground state. The field-shift [C(T,H)-C(T,0)]/C(T,0) ~ 16% at 9 T indicates a significant magnetic contribution to C(T), which is absent near $T_N$. The weak phase transition signatures suggest that thermal and transport properties may not be driven by the same mechanisms that dictate the magnetic behavior. Indeed, the energy gap for the AFM excitations, 0.83 meV, is almost negligible compared to the charge gap, 0.62 eV (**Fig.2**). The effect of the magnetic state on the transport properties is thus inconsequential near $T_N$ but more significant at low temperatures. This sharply contrasts with the behavior driven by strong couplings between the magnetic and charge gaps commonly observed in other correlated electron systems, particularly in *3d*-transition metal oxides [91,92].

In addition, the saturating resistivity at low temperatures (**Fig.5b**) is interesting, and is commonly seen in Sr$_2$IrO$_4$. This behavior, whose origin is still unclear, raises a question as to whether it is due to impurity scattering or profound mechanism(s) such as surface Dirac cones in SmB$_6$ [101].

In the following, we examine a few outstanding features of electrical and thermal transport properties that emphasize the importance of lattice degrees of freedom.

### *(a) Magnetoresistance*

The electrical resistivity is coupled to the magnetic field in a peculiar fashion and so far no available model can describe the observed magnetoresistivity shown in **Fig. 7** [50]. We focus on a representative temperature T = 35 K. Both $\rho_a$ (**H∥a**) (**Fig. 7b**) and $\rho_c$(**H∥a**) (**Fig. 7c**) exhibit an abrupt drop by ~ 60% near $\mu_o$H = 0.3 T applied along the *a*-axis, where a metamagnetic transition occurs [4,50,97,102]. These data partially track the field dependences of $M_a$(H) and $M_c$(H) shown in **Fig. 7a**, and suggest a reduction of spin scattering. However, given the small ordered moment < 0.08 $\mu_B$/Ir, a reduction of spin scattering alone certainly cannot account for such a drastic reduction in $\rho$(H). Even more strikingly, for **H ∥ c**-axis, both $\rho_a$(**H∥c**) and $\rho_c$(**H∥c**) exhibit anomalies at $\mu_o$H = 2 T and 3 T, respectively, which lead to a large overall reduction of resistivity by more than 50%. However, no anomalies corresponding to these transitions in $M_a$(H) and $M_c$(H) are discerned! In addition, dM/dH shows no slope change near $\mu_o$H = 2 and 3 T. Such behavior is clearly not due to the Lorenz force because $\rho_c$(**H∥c**) exhibits the same behavior in a configuration where both the current and **H** are parallel to the *c*-axis (**Fig. 7c**).

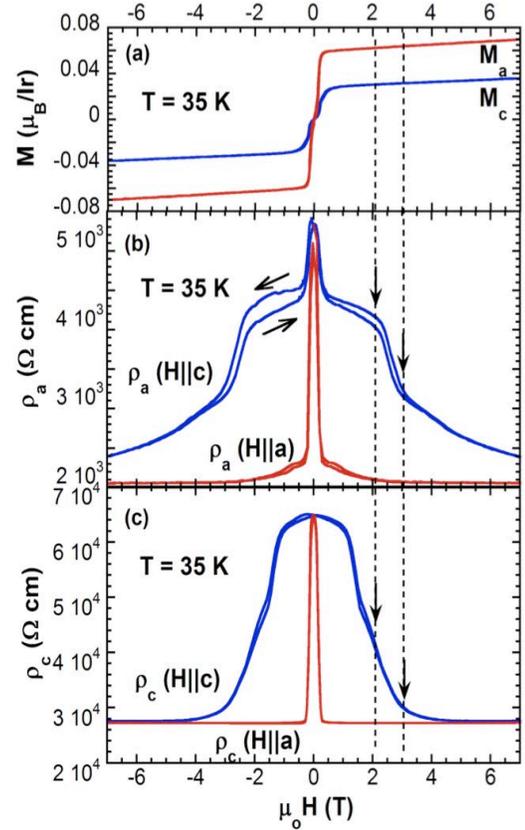

**Fig. 7.** *Sr₂IrO₄: The field dependences at T = 35 K of:* **(a)** *the magnetizations $M_a$ and $M_c$.* **(b)** *The a-axis resistivity $\rho_a$ for **H∥a** and **H∥c**.* **(c)** *The c-axis resistivity $\rho_c$ for **H∥a** and **H∥c** (Ref.50).*

An essential contributor to conventional magnetoresistivity is spin-dependent scattering; negative magnetoresistance is often a result of the reduction of spin scattering due to spin alignment with increasing magnetic field. The data in **Fig. 7** therefore raise a fundamental question: Why does the resistivity sensitively depend on the orientation of magnetic field H, but show no direct relevance to the measured magnetization when H∥*c*-axis? While no conclusive answers to the question are yet available, such varied magneto-transport behavior with temperature underscores the temperature-dependent Ir1-O2-Ir1 bond angle [4,19,50,88,98].

Moreover, such magnetotransport properties on the nanoscale are also examined using a point-contact technique [102]. Negative magnetoresistances up to 28% are discerned at modest magnetic fields (250 mT) applied within the basal plane and electric currents flowing perpendicular to the plane. The angular dependence of the magnetoresistance shows a crossover from fourfold to twofold symmetry in response to an increasing magnetic field with angular variations in resistance from 1% to 14%, which is attributed to the crystalline component of anisotropic magnetoresistance and canted moments in the basal plane. The observed anisotropic magnetoresistance is large compared to that in 3d transition metal alloys or oxides (0.1%–0.5%) and is believed to be associated with the SOI [102].

### (b) The "S"- shaped I-V characteristic and switching effect

Early studies of iridates have also uncovered a distinct feature, namely, the non-ohmic behavior [4,6]. The non-ohmic behavior exhibits current-controlled negative differential resistivity or NDR for both the $a$- and $c$-axis directions, as shown in **Fig.8**. The I-V curve near the voltage threshold $V_{th}$ (onset of negative differential resistivity) for all temperatures shows a hysteresis effect. As the current I increases further (much higher than 100 mA in this case), the ohmic behavior is restored and gives rise to an I-V curve characterized by an ''S'' shape. The S-shaped effect is categorically different from the more commonly seen ''N''-shaped effect or the Gunn effect, which is referred to as voltage-controlled NDR and attributed to electrons transferred between multienergy valleys [4,6]. It has been reported previously that the S-shaped effect is observed in some materials with a metal-insulator transition such as $CuIr_2S_{4-x}Se_x$, and is attributed to an electro-thermal effect [103]. A similar I-V characteristic has been found later in bulk single-crystal $BaIrO_3$ [6] and $Ca_3Ru_2O_7$ [6], and more recently in $VO_2$ [104]. The S-shaped I-V characteristic is restricted to the AFM insulating state. Its mechanism is still unclear, although it was suggested that the "S" effect might be related to a small band gap associated with charge density waves (CDW) [4,6]. In this case, the CDW is then pinned to the underlying lattice and slides relative to the lattice, giving rise to the NDR at $V > V_{th}$. Accordingly, one could assume a two-band model where the normal electrons and electrons in the CDW provide separate, independent channels for the conduction process [4,6]. It is clear that $\rho$ is current dependent and drastically decreases as I increases throughout the temperature range measured (see the inset of **Fig.8**) [4]. A recent study using nanoscale contacts reports a continuous reduction of the resistivity of $Sr_2IrO_4$ with increasing bias, which is characterized by a reduction in the transport activation energy by as much as 16% [90]. This behavior, which is qualitatively consistent with that reported earlier [4], is essentially attributed to changes in the Ir1-O2-Ir1 bond angle induced by the electric field [90].

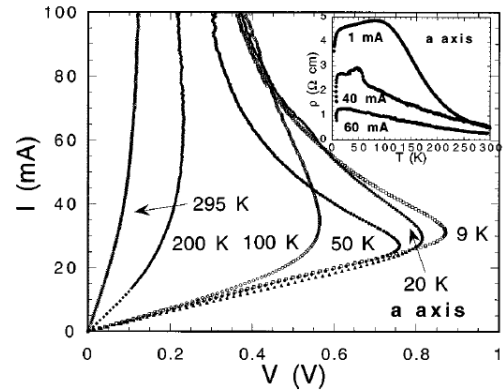

**Fig.8.** $Sr_2IrO_4$: Current I vs voltage V for various temperatures. Inset: $\rho$ along the $a$-axis vs temperature for various currents (Ref.4).

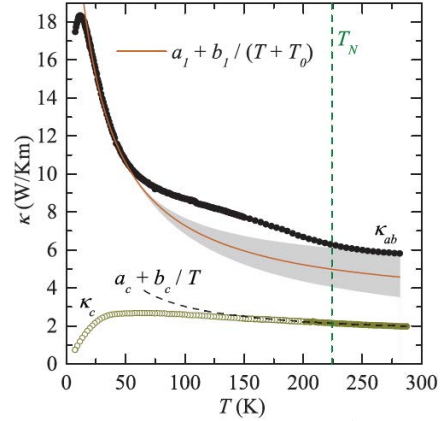

**Fig.9.** $Sr_2IrO_4$: *Heat conductivity $\kappa$ for the ab-plane ($\kappa_{ab}$) and along the c-direction ($\kappa_c$). Phononic fits to $\kappa$ are shown as solid line. The fit to the ab-direction is given by the values $a_1 = 2.8 W/Km$, $b_1 = 530 W/m$ and $T_0 = 18K$. Similarly, the fit to the c-direction (dashed line) with $a_c = 1.43 W/Km$, $b_c = 160 W/m$ (Ref.105).*

### (c) Thermal Conductivity - Pseudospin Transport

Pseudospin excitations give rise to significant thermal conductivity despite the insulating state (see **Fig.9**) [105]. The analysis of the thermal conductivity reveals a relaxation of the pseudospin excitations at low temperatures. However, the relaxation rate dramatically increases as temperature rises due to thermally activated phonon scattering. The comparison of the results with those for the cuprates with S = 1/2 spin excitations suggests a stronger coupling of the $J_{eff}$ = 1/2 pseudospin excitations to the lattice [105].

This is consistent with the other experimental observations and the underlying characteristic of the iridates that physical properties are intimately associated with the lattice owing to the strong SOI. It is noteworthy that the anomaly near $T_N$ is weak in $\kappa_{ab}$ and absent in $\kappa_c$ (**Fig.9**).

### D. *Effects of High Pressure*

A rare avoidance of metallization at high pressures in all known iridates further highlights the novelty of these materials [47,95,106,107,108,109,110]. It is commonly anticipated that an insulating state collapses and a metallic state emerges at high pressures as the unit cell shrinks and the bandwidth broadens [*e.g.* [111,112] and references therein]. In sharp contrast, $Sr_2IrO_4$ [95,109], along with other iridates (such as $Sr_3Ir_2O_7$ [47,108,109]), $BaIrO_3$ [106,107], $SrIrO_3$, $Na_2IrO_3$, $(Na_{0.10}Li_{0.90})_2IrO_3$ [113]), does not metallize at pressure up to 40 GPa (**Fig. 10**). In fact, pressure also destabilizes the ambient metallic state, resulting in an insulating state in iridates such as $SrIrO_3$ and Gd-doped $BaIrO_3$ [115]. A broad *"U-shaped"* curve often characterizes the pressure dependence of the electrical resistance of these materials -- the resistance drops initially, reaches a minimum (~20 GPa for $Sr_2IrO_4$) and then rises again as pressure increases (**Fig. 10b**). This distinct behavior defines a signature characteristic of the iridates. Moreover, an X-ray magnetic circular dichroism (XMCD) study at high pressures suggests a loss of the long-range magnetic order near 20 GPa (**Fig. 11**) but the insulating state remains (**Fig. 10a**) [95]. The nature of the insulating state above 20 GPa generates speculations that spin liquids may form in square lattices [114]. This behavior further emboldens that transport properties of $Sr_2IrO_4$ exhibit no discernable anomaly corresponding to the AFM transition at $T_N$ (= 240 K), indicating an unconventional correlation between the magnetic state and insulating gap. However, it is clear that structural distortions (such as directional bonding) play a role in the formation of an insulating gap that is much more critical than traditionally recognized. This is illustrated in many studies in which the electronic ground state of the iridates critically hinges on the bond angle Ir1-O2-Ir1of $IrO_6$ octahedra [50,55,57,115,116,117], which in turn dictates the low-energy Hamiltonian [13], thus the ground state. Indeed, the conspicuous absence of bulk superconductivity widely predicted in $Sr_2IrO_4$ is another testament to such an unconventional correlation despite structural, electronic and magnetic similarities between $Sr_2IrO_4$ and $La_2CuO_4$. This unusual character has helped revitalize discussions of Mott, Mott-Hubbard and Slater insulators, in particular, the dependence of the charge gap formation on magnetic interactions in $Sr_2IrO_4$ [52,118,119]. Remarkably, a time-resolved optical study indicates that $Sr_2IrO_4$ is a unique system in which Slater- and Mott-Hubbard-type behaviors coexist [118], which might help explain the absence of anomalies at $T_N$ in transport and thermodynamic measurements discussed above. Clearly, a better understanding of the $J_{eff}$=1/2 state and its correlation with the AFM state in $Sr_2IrO_4$ needs to be established. The unconventional correlation may ultimately change the notion of localization promoting Hund's rule magnetic moment formation vs spin liquids or spin dimers.

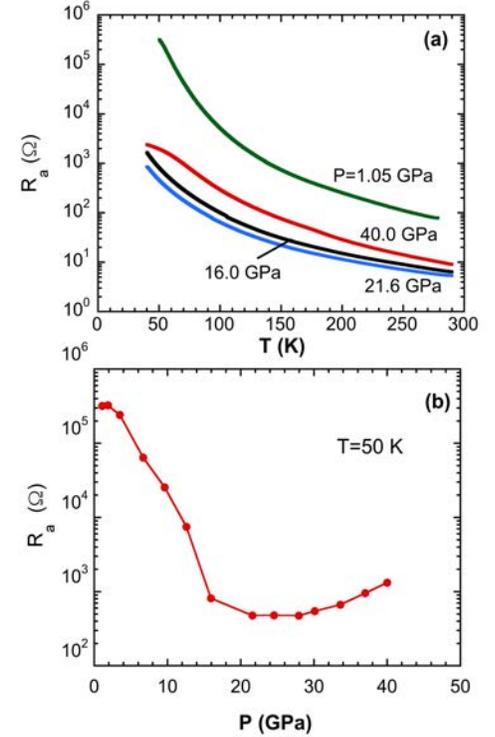

***Fig.10.*** *$Sr_2IrO_4$:* ***(a)*** *The temperature dependence of electrical resistance for the a-axis $R_a$ at a few representative pressures.* ***(b)*** *The pressure dependence of $R_a$ at a representative temperature of 50 K (Ref.95).*

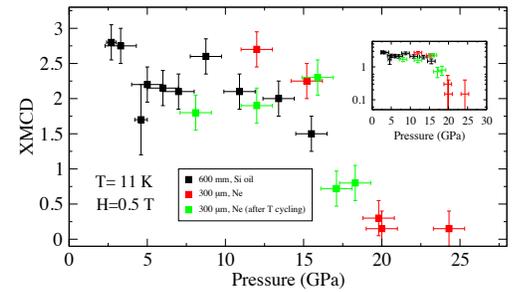

***Fig.11.*** *$Sr_2IrO_4$: The pressure dependence of XMCD at 11 K and 0.5 T. Note that the vanishing XMCD signal near 20 GPa where the electrical resistance exhibits a minimum in **Fig.10b** (Ref. 95).*

## E. *Effects of Chemical Substitution*

In stark contrast, a growing body of experimental evidence has shown that a metallic state can be readily realized via slight chemical doping, either electron (e.g. La doping [50,120]) or hole doping (e.g., K [50] or Rh [55] or Ru [53,54] doping, for either Sr or Ir or oxygen [115]) despite the sizable energy gap (~ 0.62 eV shown in **Fig.2**). Electron doping adds extra electrons to the partially filled $J_{eff}$=1/2 states, which is energetically favorable. Hole doping also adds additional charge carriers, resulting in a metallic state. However, because of the multi-orbital nature of the iridate, the mechanism of electron and hole doping in $Sr_2IrO_4$ may not be symmetrical [18]. An important distinction lies in the energy gap to the nearest $5d$ states [18,52,118,119]. However, the same effect of both electron and hole doping, as suggested by some experimental evidence, is to reduce the structural distortions or relax the buckling of $IrO_6$ octahedra, independent of the ionic radius of the dopant. As illustrated in **Fig. 12a**, a dilute doping of either $La^{3+}$ (electron doping) or $K^+$ (hole doping) ions for $Sr^{2+}$ ions leads to a larger Ir1-O2-Ir1 bond angle θ, despite the considerable differences between the ionic radii of Sr, La, and K, which are 1.18 Å, 1.03 Å and 1.38 Å, respectively. Empirical observations suggest that the reduced distortions, along with extra charge carriers, can effectively destabilize the $J_{eff}$=1/2 state, leading to a metallic state, since hopping between active $t_{2g}$ orbitals is critically linked to θ. Indeed, $\rho_a$ ($\rho_c$) is reduced by a factor of $10^8$ ($10^{10}$) at low temperatures for mere x=0.04 and 0.02 for La and K doping, respectively (**Figs. 12b, 12c** and **12d**) [50]. Another possible contribution to the reduction in the resistivity is the formation of an impurity band due to doping. Furthermore, for La doping of x = 0.04, there is a sharp downturn near 10 K, indicative of a rapid decrease in inelastic scattering (**Fig. 12c Inset**), which is similar to an anomaly observed in oxygen-depleted or effectively electron-doped $Sr_2IrO_{4-\delta}$ with $\delta = 0.04$ [115]. Other studies of La-doped $Sr_2IrO_4$ show similar effects of La doping on physical properties although behavior varies in detail [120].

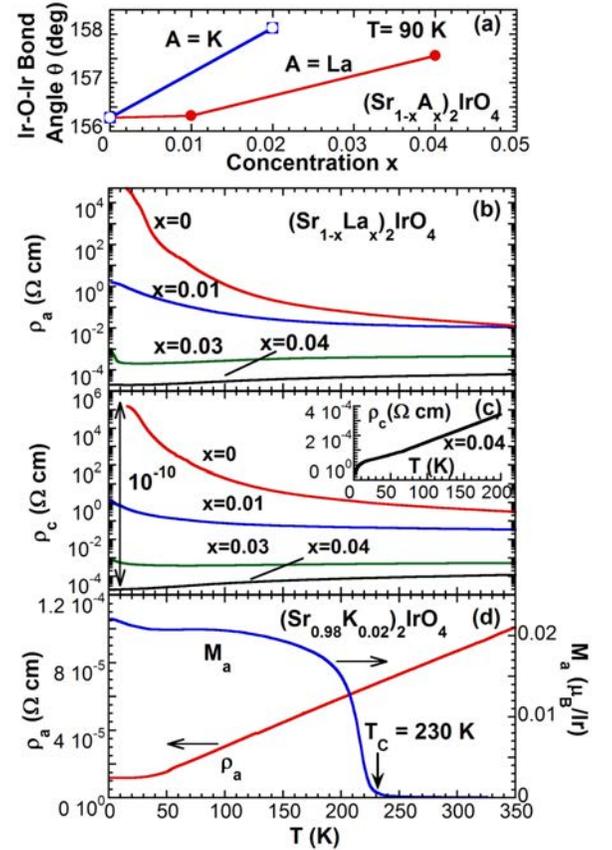

**Fig. 12.** *(a) The Ir1-O2-Ir1 bond angle θ as a function of La and K doping concentration x. The temperature dependences of (b) the a-axis resistivity $\rho_a$, and (c) the c-axis resistivity $\rho_c$ for $(Sr_{1-x}La_x)_2IrO_4$ with $0 \leq x \leq 0.04$. Inset in (c): Enlarged $\rho_c(T)$ at low T. (d) The temperature dependence of $\rho_a$ and the a-axis magnetization $M_a$ at applied field $\mu_oH = 0.1$ T (right scale) for $(Sr_{0.98}K_{0.02})_2IrO_4$ (Ref.50).*

It is noteworthy that $T_N$ decreases with La doping in $(Sr_{1-x}La_x)_2IrO_4$, and vanishes at x = 0.04, where the metallic state is fully established. In contrast, magnetic order coexists with a fully metallic state in $(Sr_{0.98}K_{0.02})_2IrO_4$, as shown in **Fig. 12d**. This comparison stresses that the occurrence of a metallic state does not necessarily demand any radical changes in the magnetic state of the iridate [50]. This observation is in accord with the absence of a resistivity anomaly near $T_N$ discussed above.

For Ru- [53,54] or Rh- [55] doped $Sr_2IrO_4$, effects of doping on magnetic and electronic properties are different in detail, but in essence Ru or Rh doping results in a metallic state that coexists with the AFM excitations. For example, for $Sr_2Ir_{1-x}Ru_xO_4$, the AFM excitations persist up to at least x = 0.77 and the maximum energy scale of the magnetic excitations at high dopings is comparable to that in undoped $Sr_2IrO_4$ [119].

Remarkably, mere 3% $Tb^{4+}$ substitution for Ir effectively suppresses $T_N$ to zero but retains the insulating state, that is, the disappearance of the AFM state accompanies no emergence of a metallic state [57]. A recent theoretical study suggests that the interaction between the magnetic moments on the impurity $Tb^{4+}$ ion and its surrounding $Ir^{4+}$ ions can be described by a "compass" model, i.e., an Ising-like interaction that favors the magnetic moments across each bond to align along the bond direction. This interaction quenches magnetic vortices near the impurities and drives a reentrant transition out of the AFM phase, leading to a complete suppression of the Néel temperature [57].

All this further highlights an unconventional correlation between the AFM and insulating states in which the magnetic transition plays a nonessential role in the formation of the charge gap in the iridate. There is experimental evidence that suggests that isoelectronic doping such as $Ca^{2+}$ doping for $Sr^{2+}$, which adds no additional charge carriers to the $J_{eff}=1/2$ state, reduces electrical resistivity in Ca doped $Sr_2IrO_4$ [51]. More recent studies on Ca and Ba doped $Sr_2IrO_4$ and $Sr_3Ir_2O_7$ single crystals show an insulator-metal transition, which is attributed to reduced structural distortions [120]. In short, a central empirical trend points to an unusually critical role of the Ir1-O2-Ir1 bond angle that mostly dictates the ground state of the iridates.

## F. Elusive Superconductivity and Odd-Parity Hidden Order

$Sr_2IrO_4$ bears key structural, electronic and magnetic features similar to those of $La_2CuO_4$ [1,2,3,4,7,9,13,16,24,23,24,25], as pointed out in the beginning of this section. In particular, the magnon dispersion in $Sr_2IrO_4$ is well-described by an AFM Heisenberg model with an effective spin one-half on a square lattice, which is similar to that in the cuprate. The magnon bandwidth in $Sr_2IrO_4$ is of 200 meV, as compared to ~ 300 meV in $La_2CuO_4$ [1,2,3,4,7,9,24]. This smaller bandwidth is consistent with other energy scales, such as hopping $t$ and U, which are uniformly smaller by approximately 50% in $Sr_2IrO_4$ than in the cuprate [22]. Largely because of these apparent similarities, a pseudo-spin-singlet d-wave superconducting phase with a critical temperature, $T_C$, approximately half of that in the superconducting cuprate is anticipated in electron-doped $Sr_2IrO_4$ whereas pseudo-spin triplet pairing may emerge in the hole-doped iridate so long as the Hund's rule coupling is not strong enough to drive the ground state to a ferromagnetic state [18,22,23,24,25,26,27,28]. A growing list of theoretical proposals has motivated extensive investigations both theoretically and experimentally in search of superconductivity in the iridates in recent years. Indeed, there is some experimental evidence signaling behavior parallel to that of the cuprates. Besides results of resonant inelastic x-ray scattering (RIXS) that indicate the similar magnon dispersion in $Sr_2IrO_4$ to that in $La_2CuO_4$ [24,121], Raman studies also reveal excitations at 0.7 eV in $Sr_2IrO_4$, about 50% smaller than 1.5 eV observed in $La_2CuO_4$ [122]. Studies of angle-resolved photoemission spectroscopy (ARPES) reveal an evolution of Fermi surface with doping (e.g., pseudogaps, Fermi arcs) that is strikingly similar to that in the cuprates and is observed in $Sr_2IrO_4$ with either electron-doping (e.g., La doping) or hole-doping (e.g., Rh doping) [119,121]. For example, the Fermi surface segments and pseudogaps for Rh doped $Sr_2IrO_4$ (hole doing) are illustrated in **Fig. 13**. It is particularly interesting that a temperature and doping dependence of Fermi arcs at low temperatures is observed with in-situ K doping in the cleaved crystal surface of $Sr_2IrO_4$. This phenomenology is strikingly similar to that of the cuprates. The similarities to the cuprates are further signified in a more recent RIXS study of La-doped $Sr_2IrO_4$ [123]. This study uncovers well-defined dispersive magnetic excitations. The dispersion is almost intact along the anti-nodal direction but exhibits significant softening along the nodal direction, similar to those in hole-doped cuprates [123].

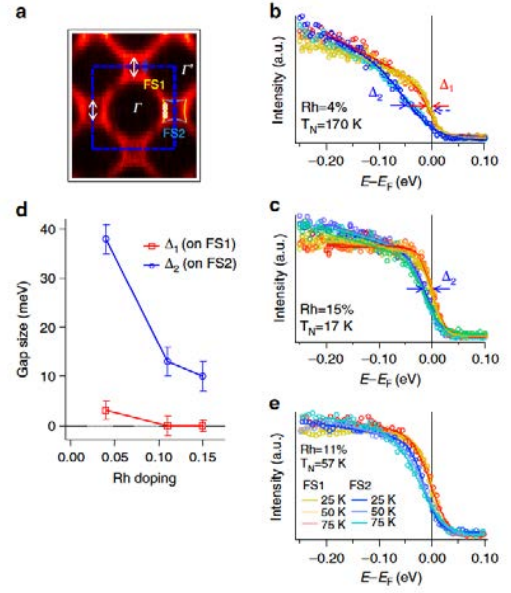

**Fig.13.** *Fermi surface (FS) segments and pseudogaps in $Sr_2Ir_{1-x}Rh_xO_4$. **(a)** The FS spectral weight for x=15%, with a hole-like Fermi pocket centered around the (π, 0) point of the unfolded (blue dashed) Brillouin zone. The FS pocket is separated into segments FS1 (yellow) and FS2 (blue), with FS1 facing Γ and FS2 facing Γ'. Q vectors (white arrows) are possible density wave nesting vectors. **(b)** and **(c)** Energy distribution curves (EDCs) from multiple locations along the FS1 and FS2 segments (yellow and blue, respectively) taken from x=4% and x=15%. The leading edges of most EDCs do not reach $E_F$, suggesting an occurrence of pseudogaps. Gap sizes are shown in (b) and (c) and compiled in **(d)**, with Δ1 labelling the gaps from FS1 and Δ2 the gaps from FS2. **(e)** EDCs from FS1 (dashed) and FS2 (solid) showing minimal temperature dependence across the magnetic phase transition for x =11% (Ref. 119).*

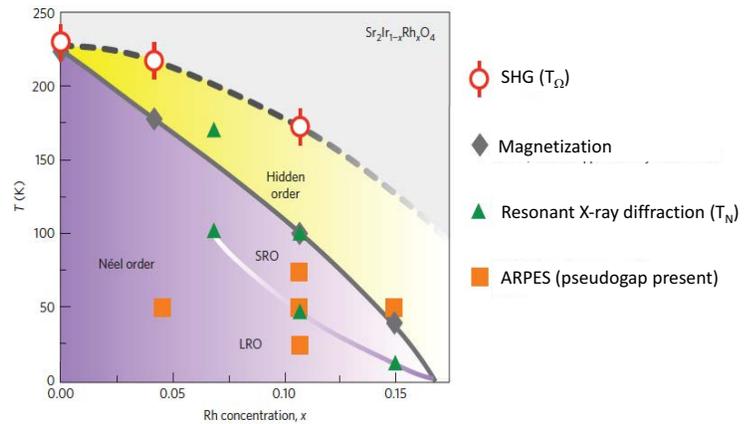

**Fig. 14.** *The phase diagram of temperature vs Rh doping for $Sr_2Ir_{1-x}Rh_xO_4$ Note the boundaries of the hidden order and the long-range (LRO) and short-range (SRO) in the AFM regions (Ref. 89).*

However, superconductivity, which is characterized by zero-resistivity and diamagnetism, remains markedly elusive, although a metallic state is a common occurrence in doped $Sr_2IrO_4$. The absence of superconductivity may be a manifestation of the particular importance of lattice properties which separates the iridates from the cuprates, in spite of all the similarities between these two classes of materials described herein and in literature. Indeed, the lattice-dependence of physical properties is much weaker in the cuprates, where the SOI is generally negligible. The spin and orbits are separated in the cuprates whereas they are strongly coupled by the SOI in the iridates. The Cooper pairs in the iridates would then be formed within pseudospin space, rather than with ordinary spins.

Interestingly, an optical second-harmonic generation (SHG) study of single-crystal $Sr_2Ir_{1-x}Rh_xO_4$ has recently revealed an odd-parity hidden order that sets in at a temperature $T_\Omega$, emerging prior to the formation of the AFM state, as shown in **Fig.14** [89] (a recent, detailed study confirms the existence of this odd-parity hidden order [124]). This order breaks both the spatial inversion and rotational symmetries of the underlying tetragonal lattice, which is expected from an electronic phase that has the symmetries of a magneto-electric loop-current order [125,126]. The onset temperature of this phase is monotonically suppressed with hole doping, although much more weakly than is the Néel temperature, revealing an extended region of the phase diagram with purely hidden order (**Fig.14**). Driving this hidden phase to its quantum critical point might open a new avenue toward realizing the widely anticipated superconducting state in $Sr_2IrO_4$ [89].

## II.2. $Sr_3Ir_2O_7$ (n=2)

The splitting between the $J_{eff} = 1/2$ and $J_{eff} = 3/2$ bands narrows as the effective dimensionality (i.e., n) increases in $Sr_{n+1}Ir_nO_{3n+1}$, and the two bands progressively broaden and contribute a finite density of states near the Fermi surface. In particular, the bandwidth W of the $J_{eff} = 1/2$ band increases from 0.48 eV for n = 1, to 0.56 eV for n = 2. and 1.01 eV for n = ∞ [8,85]. The ground state evolves with decreasing charge gap Δ as n increases, from a robust insulating state for $Sr_2IrO_4$ (n = 1) to a metallic state for $SrIrO_3$ (n = ∞). A well-defined, yet weak, insulating state exists in the case of $Sr_3Ir_2O_7$ (n = 2) [5]. Given the delicate balance between relevant interactions, $Sr_3Ir_2O_7$ is theoretically predicted to be at the border between a collinear AFM insulator and a spin-orbit Mott insulator [18,127]. Early experimental observations indicated an insulating state and a long-range magnetic order at $T_N = 285$ K with an unusual magnetization reversal below 50 K (**Figs. 15** and **16**) [5]. The value of the charge gap (180 meV) [85], much smaller than that (0.62 eV) of $Sr_2IrO_4$, but the magnetic gap is unusually large at 92 meV [128]. The borderline nature of the weak insulating state of $Sr_3Ir_2O_7$ is apparent in an ARPES study of the near-surface electronic structure, which exhibits weak metallicity evidenced by finite electronic spectral weight at the Fermi level [129].

## A.  *Critical Structural Features*

$Sr_3Ir_2O_7$ has strongly-coupled, double Ir-O layers separated from adjacent double layers along the *c*-axis by Sr-O interlayers, as shown in **Fig. 3**. The crystal structure features an orthorhombic cell with a = 5.5221 Å, b = 5.5214 Å, c = 20.9174 Å and Bbca symmetry [5] (although a recent study suggests a space group *C2/c* [130]). The $IrO_6$ octahedra are elongated along the crystallographic *c*-axis. The average Ir-O apical bond distances along the *c*-axis are 2.035 Å and 1.989 Å in the $IrO_6$ octahedra in the double layers. Like those in $Sr_2IrO_4$, the $IrO_6$ octahedra are rotated about the *c*-axis by 11° at room temperature. It is found that within a layer the rotations of the $IrO_6$ octahedra alternate in sign, forming a staggered structure, with the two layers comprising a double-layer being out of phase with one another [5].

## B.  *Magnetic Properties*

The onset of magnetic order is observed at $T_N = 285$ K in $Sr_3Ir_2O_7$ (**Fig. 15**) [5]. It is generally recognized that the magnetic ground state is AFM and closely associated with the rotation of the $IrO_6$ octahedra about the *c*-axis, which characterizes the crystal structure of both $Sr_2IrO_4$ and $Sr_3Ir_2O_7$ [1,2,3,4,5,130]. Indeed, the temperature dependence of the magnetization M(T) closely tracks the rotation of the octahedra, as reflected in the Ir-O-Ir bond angle θ (**Fig.15**). Unlike $Sr_2IrO_4$, $Sr_3Ir_2O_7$ exhibits an intriguing magnetization reversal in the *a*-axis magnetization $M_a(T)$ below $T_D = 50$ K, which marks the turning point where M(T) starts to decrease with temperature; both $T_N$ and $T_D$ can be observed only when the system is field-cooled (FC) from above $T_N$. This magnetic behavior is robust but not observed in the zero-field cooled (ZFC) magnetization, which instead remains positive and displays no anomalies that are seen in the FC magnetization [5]. It is also noteworthy that the signature of magnetic order is much weaker and not so well defined in the *c*-axis magnetization $M_c(T)$ (**Fig. 15**). It is unusual that all magnetic anomalies are absent in the case of ZFC measurements, implying that magnetostriction may occur near $T_N$ and ''lock'' up a certain magnetic configuration.

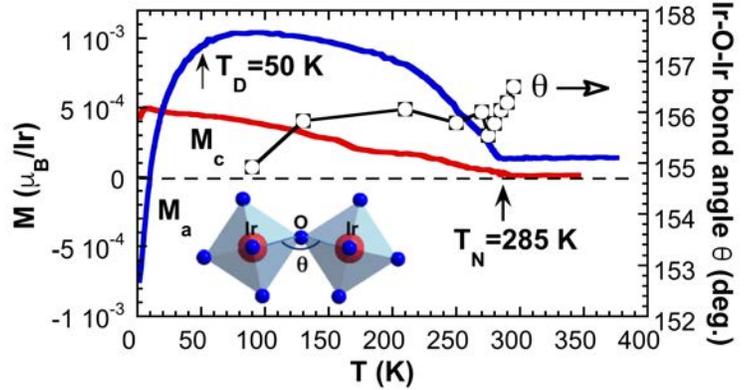

**Fig. 15.** *$Sr_3Ir_2O_7$: Temperature dependence of the magnetization for the a- and c-axis $M_a$, $M_c$, and the Ir-O-Ir bond angle θ (right scale). Note that M is measured via a field-cooled sequence; no magnetic order can be discerned in a zero-field-cooled sequence (Ref. 5).*

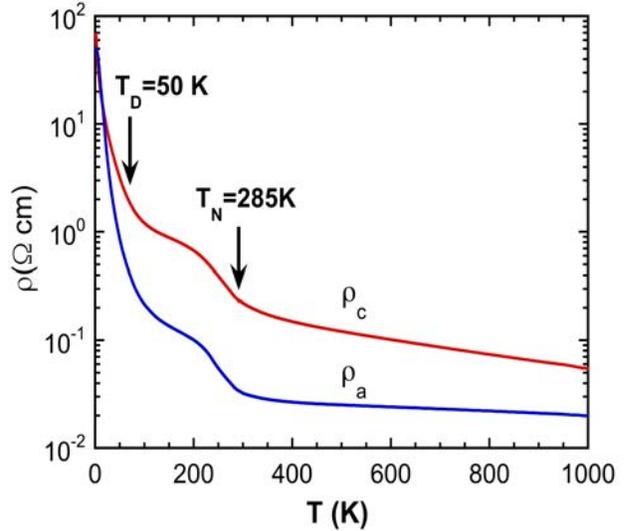

**Fig. 16.** *$Sr_3Ir_2O_7$: Temperature dependence of the a-axis resistivity $\rho_a$ and the c-axis $\rho_c$ for $1.7 < T \leq 1000$ K (Ref. 5).*

This magnetic behavior of $Sr_3Ir_2O_7$ is distinctively different from that of $Sr_2IrO_4$. A number of experimental and theoretical studies indicate that magnetic moments are aligned along the *c*-axis [131,132] but there may exist a nearly degenerate magnetic state with canted spins in the basal plane [133]. A transition to a collinear antiferromagnet via multiorbital Hubbard interactions is predicted within the mean-field approximation [18,133]. Indeed, a resonant x-ray diffraction study reveals an easy collinear antiferromagnetic structure along the *c*-axis in Sr₃Ir₂O₇, rather than within the basal plane, as observed in Sr₂IrO₄. This study further suggests a spin-flop transition as a function of the number of IrO₂ layers, due to strong competition among intra- and inter-layer bond-directional, pseudodipolar interactions [132].

The magnetic configuration of Sr₃Ir₂O₇ is highly sensitive to the lattice structure. In particular, the staggered rotation of IrO₆ octahedra between adjacent layers plays a crucial role in both insulating and magnetic states [18,133]. Sr₃Ir₂O₇ is a unique magnetic insulator, given its tiny magnetic moment [5] and is more prone to undergo a transition [18,133]. It is not surprising that any slight perturbation such as magnetic field could induce a canted spin structure, as indicated in a magnetic X-ray scattering study [134]. In addition, a more recent RIXS study reveals that the magnon dispersion is comprised of two branches well separated in energy and gapped across the entire Brillouin zone, rather than a single dominant branch. It suggests dimerization induced by the Heisenberg exchange that couples Ir ions in adjacent planes of the bilayer [135]. A better understanding of the magnetic behavior has yet to be established.

## C. *Transport Properties*

An insulating state is illustrated in the electrical resistivity $\rho(T)$ over the range $1.7 < T < 1000$ K, as shown in **Fig. 16**. The insulating behavior persists up to 1000 K in Sr₃Ir₂O₇ [5]; however, $\rho$ is at least four orders of magnitude smaller than that of Sr₂IrO₄ (**Fig. 5b**). Both $\rho_a$ and $\rho_c$ increase slowly with temperature decreasing from 1000 to 300 K, but then rise rapidly in the vicinity of $T_N$ and $T_D$, demonstrating a coupling between magnetic and transport properties (although the magnetoresistivity is negligible at high temperatures and becomes sizable only below 50 K [5].) Nevertheless, the transport behavior of $Sr_3Ir_2O_7$ contrasts with that of $Sr_2IrO_4$, where such a correlation is absent. The differing behavior may be due in part to the difference in the charge gap $\Delta$ between $Sr_3Ir_2O_7$ ($\Delta \sim 0.18$ eV) and $Sr_2IrO_4$ ($\Delta \sim 0.62$ eV) and, thus, the relative effect of the AFM excitation energy on the charge gap. A Raman study also suggests different influence of frustrating exchange interactions on $Sr_2IrO_4$ and $Sr_3Ir_2O_7$ [100]. Finally, like those of $Sr_2IrO_4$, transport properties of $Sr_3Ir_2O_7$ can be tuned electrically [136].

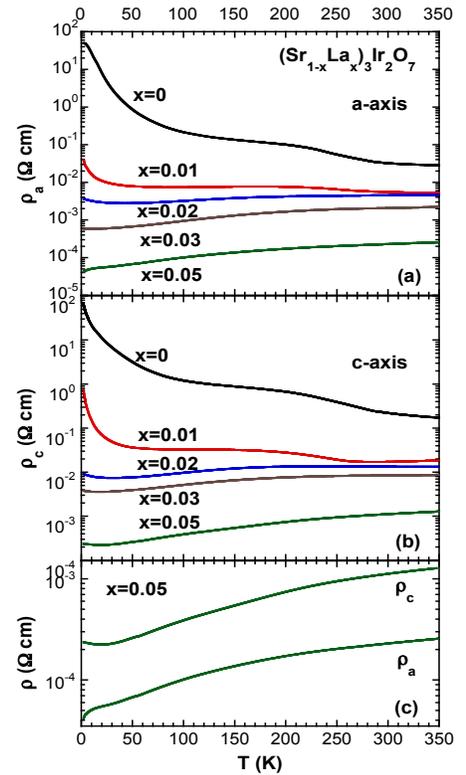

**Fig. 17.** *Temperature dependence of (a) the a-axis resistivity $\rho_a$; and (b) the c-axis resistivity $\rho_c$ for (Sr₁₋ₓLaₓ)₃Ir₂O₇. (c) $\rho_a$ and $\rho_c$ for x = 0.05 (Ref.137).*

## D. *Effects of Chemical Substitution*

The effects of chemical doping in $Sr_3Ir_2O_7$ are similar to those in $Sr_2IrO_4$ discussed in **II.1.E**. For example, 5% La doping readily precipitates a metallic state reflected in the electrical resistivity $\rho$ of single-crystal $(Sr_{1-x}La_x)_3Ir_2O_7$, as shown in **Fig. 17** [137]. The *a*-axis resistivity $\rho_a$ (the c-axis resistivity $\rho_c$) is reduced by as much as a factor of $10^6$ ($10^5$) at low temperatures as x evolves from 0 to 0.05, (see **Figs. 17a** and **17b**). For x = 0.05, there is a sharp downturn in $\rho_a$ near 10 K, indicative of a rapid decrease in inelastic scattering (**Fig. 17c**). Such low-temperature behavior is observed in oxygen-depleted $Sr_2IrO_{4-\delta}$ with $\delta = 0.04$, and La-doped $Sr_2IrO_4$, as discussed in **II.1.E**. La doping not only adds electrons to states but also significantly increases the Ir-O-Ir bond angle $\theta$, which is more energetically favorable for electron hopping and superexchange interactions.

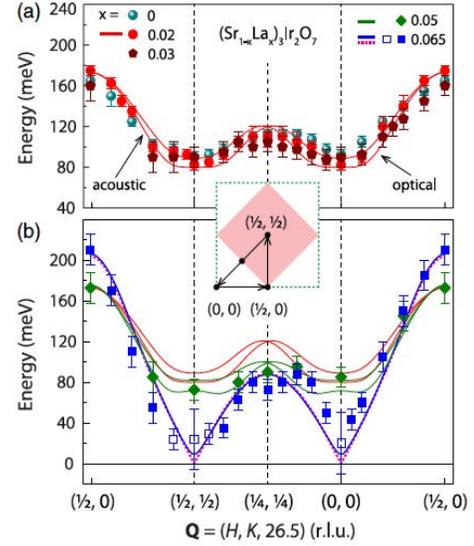

*Fig.18*. *Doping-dependent magnon dispersions for $(Sr_{1-x}La_x)_3Ir_2O_7$: (a) x=0, 0.02, and 0.03 and (b) x=0.05 and 0.065. The dispersions for $x \leq 0.05$ and the blue solid squares of x = 0.065 are obtained by selecting the peak positions of the magnetic excitations. The blue open squares of x =0.065 are extracted from fitting of the magnetic excitations. The solid curves are fits to the dispersions for x=0.02, 0.05, and 0.065 using the bilayer model, which includes an acoustic and an optical branch. The pink dashed curve is the fitting of the dispersion for 0.065 using the $J - J2 - J3$ model (Ref.138).*

It should be stressed that the occurrence of the metallic state in this system is not accompanied by a complete disappearance of the magnetic order, although it is significantly weakened. A study of $(Sr_{1-x}La_x)_3Ir_2O_7$ utilizing resonant elastic and inelastic x-ray scattering at the Ir-$L_3$ edge reveals that with increasing doping x, the three-dimensional long-range AFM order is gradually suppressed and evolves into a three-dimensional short-range order across the insulator-to-metal transition from x = 0 to x = 0.05, which is then followed by a transition to two-dimensional short range order between x = 0.05 and x = 0.065. Because of the interactions between the $J_{eff}$=1/2 pseudospins and the emergent itinerant electrons, magnetic excitations undergo damping, anisotropic softening, and gap collapse, accompanied by spin-orbit excitons that are weakly dependent on doping (**Fig.18**) [138]. It is also suggested that electron doping suppresses the magnetic anisotropy and interlayer couplings and drives $(Sr_{1-x}La_x)_3Ir_2O_7$ into a correlated metallic state with two-dimensional short-range AFM order. Strong AFM fluctuations of the $J_{eff}$ = 1/2 moments persist deep in this correlated metallic state, with the magnetic gap strongly suppressed [138]. More intriguing, a subtle charge-density-wave-like Fermi surface instability is observed in the metallic region of $(Sr_{1-x}La_x)_3Ir_2O_7$ near 200 K which shows resemblance to that observed in cuprates [139]. The absence of any signatures of a new spatial periodicity below 200 K seems to suggest an unconventional and possibly short-ranged density wave order [139].

### E. *Effects of High Pressure*

$Sr_3Ir_2O_7$ responds to pressure in a fashion similar to that of $Sr_2IrO_4$ at lower pressures as discussed above. An

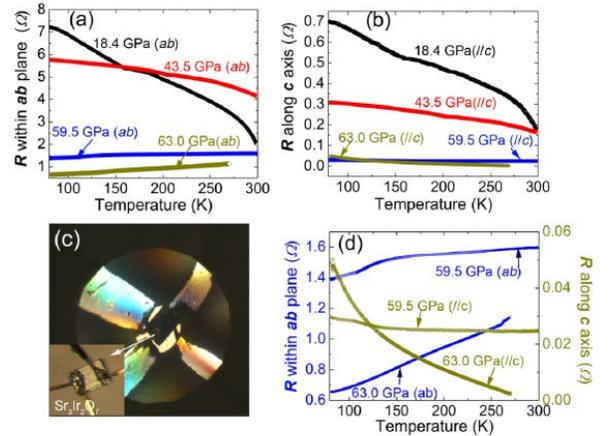

**Fig. 19.** *$Sr_3Ir_2O_7$: Electrical resistance of single-crystal $Sr_3Ir_2O_7$ at high pressure as a function of temperature for (a) the basal plane and (b) the c-axis. (c) Four gold electric leads and the sample loaded into a symmetric diamond anvil cell. The inset shows two of the leads attached to the top of the single crystal, and the other two attached to the bottom. (d) Temperature dependence of the electrical resistances at 59.5 and 63.0 GPa. Note the metallic behavior in the basal plane, and the insulating behavior along the c-axis (Ref.47).*

early pressure study reveals an abrupt change in electrical resistivity along with an apparent second-order structural change near 13 GPa [108]. A more recent study using both RIXS and electrical resistivity indicates that $Sr_3Ir_2O_7$ becomes a confined metal at 59.5 GPa, featuring a metallic state in the basal plane but an insulating behavior along the *c*-axis (**Fig.19**) [47]. This novel insulator-metal transition is attributed to a possible first-order structural change at nearby pressures [47]. The study further suggests that the structural transition above 54 GPa is likely triggered by a saturated $IrO_6$ octahedron rotation, which inevitably leads to changes in the band structure in $Sr_3Ir_2O_7$ [47]. As recognized now, the pressure-induced metallic state does not commonly occur, but when it does, it has an extraordinary nature, such as in this case.

## II.3. SrIrO₃ and its Derivative (n=∞)

The effects of the SOI and U are largely local and remain essentially unchanged throughout the entire series $Sr_{n+1}Ir_nO_{3n+1}$, and it is now known that the ground state evolves with increasing dimensionality (i.e., n), from a robust AFM insulating state for $Sr_2IrO_4$ (n = 1) to a paramagnetic, semimetallic state for $SrIrO_3$ (n =∞) [8,45,48]. The semimetallic state of $SrIrO_3$ has been of great interest both theoretically and experimentally, despite the presence of strong SOI [8,45,140,141,142,143,144,145,146,147,148,149,150]. Initial work indicated that a strong SOI reduces the threshold of U for a metal-insulator transition [140,141]. A more recent theoretical study finds that an even larger critical U is required for a metal-insulator transition to occur in $SrIrO_3$, due to a combined effect of the lattice structure, strong SOI and the protected line of Dirac nodes in the $J_{eff} = 1/2$ bands near the Fermi level [142]. In essence, small hole and electron pockets with low densities of states, which are present in $SrIrO_3$, render U less effective in driving a magnetic insulating state because of protected line node [142]. We stress that tuning the relative strength of the SOI and U is highly effective in changing the nature of the ground state in the iridates. The rare occurrence of a semimetallic state in $SrIrO_3$ therefore provides a unique opportunity to closely examine the intricate interplay of the SOI, U and lattice degrees of freedom, as well as the correlation between the AFM state and the metal-insulator transition in iridates. Indeed, tuning the relative strength of the SOI and U effectively changes the ground state in the iridates.

A large number of experimental studies of the orthorhombic perovskite $SrIrO_3$ have been conducted in recent years [8,45,140,141,142,143,144,145,146,147,148,149,150]. However, the bulk single-crystal $SrIrO_3$ forms only at high pressures and high temperatures [48], and almost all studies of this system are limited to samples in thin-film or polycrystalline form; little work on bulk single-crystal samples of this system has been done; critical information concerning magnetic properties and their correlation with the electronic state, corresponding anisotropies, etc., is still lacking. However, a recent study on bulk single-crystals of Ir-deficient, orthorhombic perovskite $Sr_{0.94}Ir_{0.77}O_{2.68}$ offers some insights [151]. In essence, $Sr_{0.94}Ir_{0.77}O_{2.68}$ retains the very same crystal structure as stoichiometric $SrIrO_3$, albeit with a rotation of $IrO_6$ octahedra within the basal plane by about 9.54° [151], which is a structural signature for all layered perovskite iridates [1,2,3,4,5,88,130].

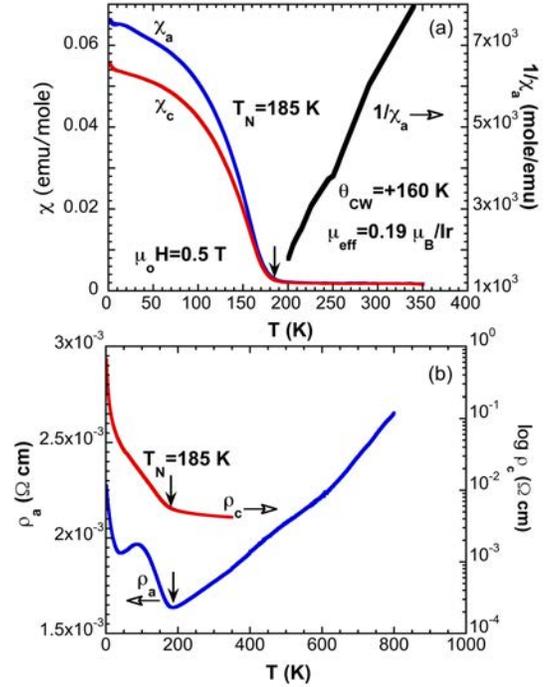

**Fig.20.** $Sr_{0.94}Ir_{0.77}O_{2.68}$: The temperature dependence of the magnetic susceptibility $\chi$ for a-axis $\chi_a$ and c-axis $\chi_c$, and $\chi_a^{-1}$ (right scale) at $\mu_o H = 0.5$ T, and **(b)** the a-axis resistivity $\rho_a$ up to 800 K, and the c-axis resistivity $\rho_c$ (right scale) (Ref.151).

$Sr_{0.94}Ir_{0.77}O_{2.68}$ exhibits sharp, simultaneous AFM and metal-insulator transitions at $T_N = 185$ K with a charge gap of 0.027 eV, sharply contrasting stoichiometric $SrIrO_3$, which is paramagnetic and semimetallic (**Fig.20**). Recalling the rotation of $IrO_6$ octahedra in $Sr_2IrO_4$ and $Sr_3Ir_2O_7$ [1,2,3,4,5] is critical in determining the AFM ground state [13], one suspects that the emerging AFM state in $Sr_{0.94}Ir_{0.77}O_{2.68}$ might be in part a result of the increased in-plane rotation of $IrO_6$ octahedra, $9.54^o$ (compared to $8.75^o$ for $SrIrO_3$ [48]). Alternatively, the absence of proper glide symmetry in $Sr_{0.94}Ir_{0.77}O_{2.68}$ weakens the nodal line protection, thus U becomes more effective, leading to the AFM state [143]. The transport properties feature an extended regime of linear-temperature basal-plane resistivity between 185 K and 800 K, and an abrupt sign change in the Hall resistivity below 40 K (rather than $T_N$=185K) that signals a transition from hole-like to electron-like behavior with decreasing temperature [151]. These studies of $SrIrO_3$ underscore the delicacy of a metallic state that is in close proximity to an AFM insulating state, and a ground state highly sensitive to slight lattice defects. The simultaneous AFM and metal-insulator transitions illustrate a direct correlation between the AFM transition and charge gap in $SrIrO_3$, which is notably absent in $Sr_2IrO_4$.

The contrasting ground states observed in isostructural $Sr_{0.94}Ir_{0.77}O_{2.68}$ and $SrIrO_3$ with space group Pbnm highlight the ultra-high sensitivity of SOI-coupled iridates to the lattice which, along with the delicate interplay of the SOI and U, need to be adequately addressed both experimentally and theoretically.

## III. Magnetism of Honeycomb Lattices and Other Geometrically Frustrated Iridates

The interest in frustrated iridates received a major boost when a theoretical analysis [13] showed that the oxygen-mediated superexchange processes between the $Ir^{4+}$ moments in the honeycomb iridates $Na_2IrO_3$ and $Li_2IrO_3$ can related to the celebrated Kitaev model applied to the $J_{eff} = 1/2$ degrees of freedom. The Kitaev model can be solved exactly, and its ground state is an exotic magnetically disordered quantum "spin liquid" [152]. However, the long-sought spin-liquid state has remained elusive. That both honeycomb lattices $Na_2IrO_3$ and $Li_2IrO_3$ (including β and γ phases) are magnetically ordered suggests that the Heisenberg interaction is still sizable. In addition, trigonal crystal fields also compete with the Kitaev interaction.

### III.1. Two-Dimensional Honeycomb Lattices

## A. Na₂IrO₃ and Li₂IrO₃

The honeycomb lattices feature $IrO_6$ octahedra that are edge-sharing with $90^o$ Ir-O-Ir bonds. The magnetic exchange is anisotropically bond-dependent. Such a bond-dependent interaction gives rise to strong frustration when Ir ions are placed on a honeycomb lattice, and would seem to favor a Kitaev spin liquid [13,152,152]. Theoretical treatments of the honeycomb lattices $Na_2IrO_3$ and $Li_2IrO_3$ have inspired a large body of experimental work that anticipates Kitaev physics [13,15,31,32,33,34,38,65,67,153,154,155,156,157,158]. If individual spins at the sites of a honeycomb lattice are restricted to align along any one of the 3 bond directions (six degrees of freedom for "up" and "down" spins), the Kitaev model predicts a quantum spin-liquid ground state [152]. This novel state features short-range correlations, and the spin degrees of freedom fractionalize into Majorana fermions and a $Z_2$ gauge field. The honeycomb iridates are often described in terms of competing Heisenberg and Kitaev interactions; the former favors an AFM state, and the latter a spin liquid state. However, no experimental confirmation of the spin liquid state has been reported, and it is experimentally established that all known honeycomb iridates order magnetically [31,32,33,65,67,153,154,155,156,157,158,159,160].

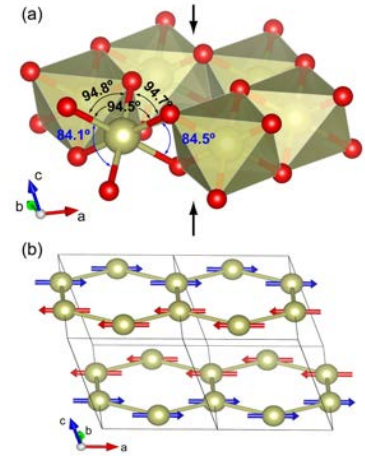

**Fig.21.** *Na₂IrO₃:* **(a)** *Local structure within the basal plane. The compression of $IrO_6$ octahedron along the stacking marked by the black arrows leads to the decrease of O-Ir-O bond angles across the shared edges.* **(b)** *The zigzag order that is consistent with the symmetry associated with observed magnetic reflections. The Ir moments between honeycomb layers are antiferromagnetically coupled (Ref. 33).*

$Na_2IrO_3$ exhibits a peculiar "zig-zag" magnetic order at $T_N$=15 K with a Mott gap of ~ 0.42 eV [160]. The magnetic order was first reported in Refs. [30,31] and later confirmed by neutron diffraction and other studies that indicate that $Na_2IrO_3$ orders magnetically below 18.1(2) K with $Ir^{4+}$ ions forming zig-zag spin chains within the layered honeycomb network with an ordered moment of 0.22(1)$\mu_B$/Ir (**Fig.21**) [33]. Inelastic neutron scattering on polycrystalline samples offers some insights into the magnetic state [32], whereas RIXS studies characterize a branch of magnetic excitations at high-energies near 30 meV [159]. The magnetic state of $\alpha-Li_2IrO_3$ is not as well-characterized as that of $Na_2IrO_3$, partly due to the lack of large single-crystal samples needed for more definitive magnetic studies. $\alpha-Li_2IrO_3$ was initially reported to feature a paramagnetic phase [161], and then AFM order at $T_N$=15 K [153]; more recently, an incommensurate magnetic order with Ir magnetic moments counter-rotating on nearest-neighbor sites was reported [63].

Indeed, the magnetic ground state of $Na_2IrO_3$ and $\alpha-Li_2IrO_3$ appear not to be related. Studies of single-crystal $(Na_{1-x}Li_x)_2IrO_3$ indicate that, as x is tuned, the lattice parameters evolve monotonically from Na to Li, and retain the honeycomb structure of the $Ir^{4+}$ planes and Mott insulating state for all x. However, there is a non-monotonic, dramatic change in $T_N$ with x, in which $T_N$ initially decreases from 18 K at x = 0 to 1.2 K at x = 0.70, before it rises to 7 K at x = 0.90 (see **Fig. 22**) [154]. Indeed, the corresponding frustration parameter $\theta_{CW}/T_N$ ($\theta_{CW}$ is the Curie-Weiss temperature) peaks at x = 0.70 at a value of 31.6, which suggests strongly enhanced frustration. The complicated evolution of the magnetic behavior clearly demonstrates that the magnetic ground states attained at x = 0 and 1 are indeed not related linearly [154] as had been previously suggested [36]. X-ray structure data also show that the $Ir^{4+}$ honeycomb lattice is minimally distorted at x ≈ 0.7, where the lowest $T_N$ and highest frustration parameter are observed. In addition, the high-temperature anisotropy in the magnetic susceptibility is simultaneously reversed and enhanced upon Li doping. Another study of $(Na_{1-x}Li_x)_2IrO_3$ suggests a phase separation in the range 0.25 < x < 0.6 [162], which calls for more investigations. Nevertheless, the evolution of structural, thermodynamic, and magnetic properties of $(Na_{1-x}Li_x)_2IrO_3$ with x suggests two possible tuning parameters for the phase transition: The crystal field splitting and the anisotropy of the distortion of the honeycomb layers, both of which change sign from the Na to Li compounds, likely compete in such a manner that explains the magnetic ordering. A relevant study of Ti-doped honeycomb lattices reveals the Curie-Weiss temperature decreases with increasing x in $Na_2(Ir_{1-x}Ti_x)O_3$, but remains essentially unchanged in $Li_2(Ir_{1-x}Ti_x)O_3$ [163], which further highlights the distinct differences between the magnetic ground states of $Na_2IrO_3$ and $Li_2IrO_3$.

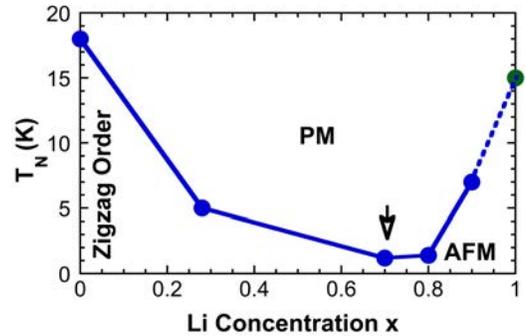

**Fig.22.** *$(Na_{1-x}Li_x)_2IrO_3$: The Néel temperature $T_N$ a function of x. Note that the lowest $T_N$ is 1.2 K at x=0.7 (Ref.154).*

There are many theoretical proposals for interactions supplementary to the Kitaev model which would cause magnetic ordering, including additional exchange processes, strong trigonal fields, or weak coupling instabilities. A consensus is yet to be reached. It is recently suggested that the Kitaev spin liquid on the honeycomb lattice is extremely fragile against the second-nearest-neighbor Kitaev coupling, which is a dominant perturbation beyond the nearest-neighbor model in $Na_2IrO_3$ [158]. It is thought that this coupling accounts for the zig-zag ordering observed in $Na_2IrO_3$ [158].

Finally, it is worth mentioning that chemical doping cannot induce a metallic state in the honeycomb iridates, despite extensive experimental efforts [164]. This behavior contrasts with that in the perovskite iridates, in which chemical doping can readily prompt a metallic state, as discussed in **Section II**, may be due to the multi-orbital nature of the honeycomb lattices (It is noted that Na or Li ions are out of plane in the unit cell, thus chemical substitution for Na or Li might only lead to a small impurity potential). On the other hand, like the perovskite iridates, the honeycomb iridates do not metallize up to 40 GPa [165].

**B. *Ru-based Honeycomb Lattices $Na_2RuO_3$, $Li_2RuO_3$, and $\alpha$-RuCl$_3$***

The investigations of the honeycomb iridates have also spread to their ruthenate counterparts, $Na_2RuO_3$ and $Li_2RuO_3$, and, more recently, Ru based $\alpha$-RuCl$_3$ (with $Ru^{3+}(4d^5)$). Both of the ruthenates feature $Ru^{4+}(4d^4)$ ions and a weaker or "intermediate strength" SOI (~ 0.16 eV, compared to ~ 0.4 eV for Ir ions) (see **Table 2**). The different d-shell filling and contrasting hierarchy of energy scales between the ruthenates and iridates provide a unique opportunity to gain a deeper understanding of the fundamental problem of interacting electrons on the honeycomb lattices. The magnetism of $Ru^{4+}$ ions as well as other "heavy $d^4$ ions" (such as $Rh^{5+}(4d^4)$, $Re^{3+}(5d^4)$, $Os^{4+}(5d^4)$ and $Ir^{5+}(5d^4)$) is interesting in their own right, as emphasized recently [166,167,168,169,170,171]. Materials with heavy $d^4$ ions tend to adopt a low-spin state because larger crystal fields often overpower the Hund's rule coupling. On the other hand, SOI with the intermediate strength may still be strong enough to impose a competing singlet ground state or a total angular momentum $J_{eff} = 0$ state. Novel magnetic states may thus emerge when the singlet-triplet splitting (0.05-0.20 eV) becomes comparable to exchange interactions (0.05-0.10 eV) and/or non-cubic crystal fields [166].

Both $Na_2RuO_3$ and $Li_2RuO_3$ adopt a honeycomb structure. $Na_2RuO_3$ features a nearly ideal honeycomb lattice with space group *C2/m*, and orders antiferromagnetically below 30 K [170]. On the other hand, single-crystal $Li_2RuO_3$ adopts a less ideal honeycomb lattice with either *C2/m* or more distorted *P2₁/m* space group below 300 K; both phases exhibit a well-defined, though different, magnetic state. It is interesting that the magnetic state in single-crystal $Li_2RuO_3$ contrasts with the nonmagnetic state or singlet ground state due to dimerization observed in polycrystalline $Li_2RuO_3$ [172]. Careful examinations of structural details reveal that these honeycomb lattices feature two bond distances (long and short, $L_l$ and $L_s$, respectively), which, to a large extent, dictate the ground state of the honeycomb lattice $Na_2RuO_3$ and $Li_2RuO_3$ (see **Fig.23**), as discussed below.

A study of these materials [170] show that the magnetic ordering temperature systematically decreases with increasing $(L_l-L_s)/L_s$ and eventually vanishes at a critical value where dimerization emerges, which leads to the singlet ground state observed in polycrystalline $Li_2RuO_3$. A phase diagram generated in this study uncovers a direct correlation between the ground state and basal-plane distortions (lattice-tuned magnetism) in the honeycomb ruthenates, as shown in **Fig. 23** [170]. In addition, a few other observations are remarkable: (1) Both $Li_2RuO_3$ and $Li_2IrO_3$ are more structurally distorted and behave with more complexities than their Na counterparts. (2) Although the SOI is expected to impose a singlet $J_{eff} = 0$ state for $Ru^{4+}(4d^4)$ ions (and a $J_{eff} = 1/2$ state for $Ir^{4+}(5d^5)$ ions), the observed magnetic states in the honeycomb ruthenates as in many other ruthenates [19] indicate that the SOI is not sufficient to induce a $J_{eff} = 0$ state. (3) All honeycomb ruthenates and iridates magnetically order in a similar temperature range despite the different role of the SOI in them [170].

A Ru-based honeycomb lattice, $\alpha$-$RuCl_3$, with $Ru^{3+}(4d^5)$ (rather than the commonplace $Ru^{4+}$ state for ruthenates) [173,174] has recently drawn a great deal of attention as a candidate for the spin liquid state [175,176,177,178,179]. $\alpha$-$RuCl_3$, with space group $P3_112$ (C2/m was also reported), supports a Mott state with an AFM order below $T_N = 7$ K [179] ($T_N$ is sample dependent in early reports; it is now recognized that $T_N = 7$ K is intrinsic). The magnetic structure is determined to be zigzag, similar to that of $Na_2IrO_3$ [179,179]; but $\alpha$-$RuCl_3$ adopts a more ideal honeycomb lattice without the distortions found in $Na_2IrO_3$. This simplified structure, combined with the weaker neutron absorption cross-section of Ru compared to Ir, favors $\alpha$-$RuCl_3$ for further experimental studies of Kitaev physics. On the other hand, the weaker SOI in the 4d shell might result in a large overlap between the $J_{eff} = 1/2$ and $J_{eff} = 3/2$ bands, thereby reducing the signature of the $J_{eff}$ states in experimental data. This issue is addressed in recent studies [18,177,179].

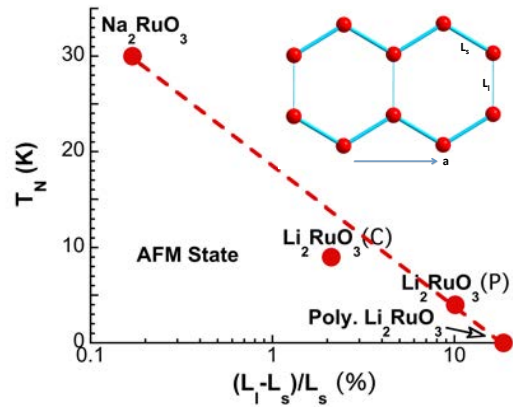

**Fig.23.** *A phase diagram for honeycomb ruthenates: The Néel temperature $T_N$ as a function of the bond distance ratio $(L_l-L_s)/L_s$ (%) for all honeycomb ruthenates. Inset: a schematic of the honeycomb lattice featuring long bond $L_l$ and short bond $L_s$ (Ref. 170).*

### III.2 $\beta$-$Li_2IrO_3$ and $\gamma$-$Li_2IrO_3$

Two derivatives of the two-dimensional honeycomb lattices discussed above are the hyper-honeycomb $Li_2IrO_3$ and stripy-honeycomb $Li_2IrO_3$, formally termed, β-$Li_2IrO_3$ and γ-$Li_2IrO_3$, respectively. They are a result of strong, mainly trigonal and monoclinic distortions of networks of edge-shared octahedra similar to those found in $Na_2IrO_3$ and α-$Li_2IrO_3$. With a $J_{eff}$ = 1/2 state, both β-$Li_2IrO_3$ and γ-$Li_2IrO_3$ antiferromagnetically order at 38 K into incommensurate, counter-rotating spirals. But unlike the in-plane moments in α-$Li_2IrO_3$, the β- and γ-phase moments are non-coplanar, forming three-dimensional honeycomb lattices [65,67]. There are some subtle differences in magnetic moments, but the two distorted honeycomb lattices are strikingly similar in terms of incommensurate propagation vectors ([0.57,0,0]) [64,66], which implies a possible common mechanism drives the ground state. A burgeoning list of theoretical proposals emphasize a combined effect of the ferromagnetic Kitaev limit, structural distortions and exchange interactions between nearest neighbors (or even next-nearest neighbors), which may account for the stability of the incommensurate order on the three-dimensional honeycomb lattice [158,180,181,182,183,184].

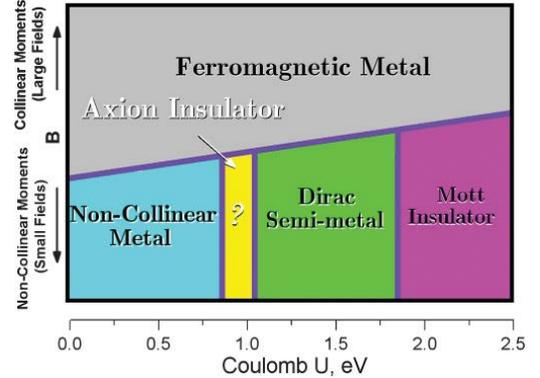

**Fig. 24.** Pyrochlore iridates: The predicted phase diagram. The horizontal axis corresponds to U whereas the vertical axis represents external magnetic field, which can trigger a transition out of the noncollinear "all-in/all-out" ground state that includes several electronic phases (Ref.29).

### III.3. Hyperkagome $Na_4Ir_3O_8$ and Pyrochlore Iridates

A large number of theoretical and experimental studies on other frustrated iridates, such as the hyper-kagome $Na_4Ir_3O_8$ [20,185,186] and the pyrochlore iridates [17,71,72,187], have been conducted [17].

$Na_4Ir_3O_8$ with a frustrated hyperkagome lattice was first reported in 2007 [20]. It features magnetic and thermal properties appropriate for a spin-liquid state (e.g., no long-range magnetic order above 2 K and no magnetic field dependence of the magnetization and heat capacity) [20]. More recent studies confirmed the absence of long-range magnetic order down to T=75 mK [185]. This work [20] has helped generate a great deal of interest in geometrically frustrated iridates.

The pyrochlore iridates, $R_2Ir_2O_7$ (R = Y, rare earth ion or Bi or Pb), are mostly magnetic insulators, except for $Pr_2Ir_2O_{7-\delta}$, which exhibits correlated metallic behavior [188,189,190,191], $Bi_2Ir_2O_7$ and $Pb_2Ir_2O_7$ [68,69,70], which also present a metallic ground state. Many of these iridates have been intensively studied as potential platforms for exotic states such as spin liquids, Weyl semimetals, axion insulators, topological insulators, etc. [17,29]. A recent study using scanning microwave imaging reveals one-dimensional metallic channels along all-in-all-out (AIAO)/all-out-all-in magnetic domain walls [187]. A rich phase diagram has been predicated for the pyrochlore iridates (see **Fig.24**) [29]. It is established that the ground state of these materials sensitively depends on the relative strengths of competing SOI and U, and a hybridization interaction that is controlled by the Ir-O-Ir bond angle [17,29,191,192]; and therefore, small perturbations can easily tip the balance between the competing energies and ground states [17,191]. This is perfectly illustrated in a phase diagram presented in Ref. **[191]**, in which the ground state of $R_2Ir_2O_7$ (R = rare earth ion) sensitively depends on the ionic radius of the rare earth ion [17,191]. However, the nature of the magnetic state is yet to be conclusively determined, in part due to experimental challenge of synthesizing large, high-quality single crystals of the pyrochlore iridates.

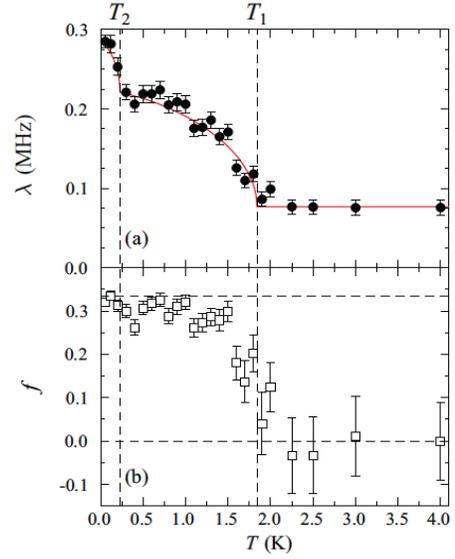

**Fig. 25.** $Bi_2Ir_2O_7$: **(a)** μ-spin relaxation rate λ. The solid line is a guide to the eye. **(b)** Fractional amplitude of the nonrelaxing asymmetry, f. Dashed lines show transition temperatures and limiting values of f (Ref.70).

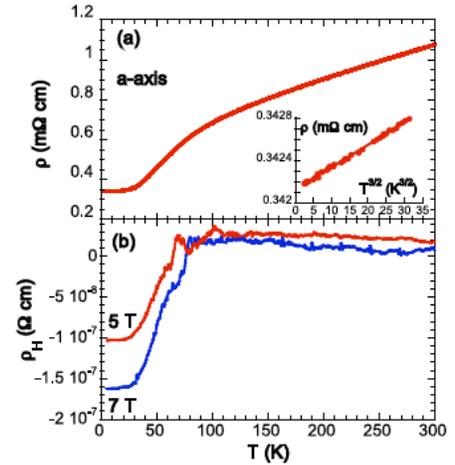

**Fig. 26.** $Bi_2Ir_2O_7$: The temperature dependence of **(a)** the a-axis electrical resistivity ρ, and **(b)** the Hall resistivity at 3 T and 5 T.
Inset: the a-axis ρ versus $T^{3/2}$ for 1.7 K< T <10 K (Ref.69).

A relevant, but less studied pyrochlore iridate is $Bi_2Ir_2O_7$ (where the $Bi^{3+}$ ion substitutes for a rare-earth ion) [68]. Investigations of single-crystal $Bi_2Ir_2O_7$ indicate a significantly enhanced hybridization between the Bi-6s/6p and Ir-5d electrons, which overpowers the SOI and U, and drives the material into a metallic state with the Fermi energy residing near a sharp peak in the density of states, despite the large $Z$ (thus SOI) for both Bi and Ir [69,70,193]. Muon spin relaxation (μSR) measurements show that $Bi_2Ir_2O_7$ undergoes a bulk magnetic transition at 1.84(3) K (**Fig. 25a**) [70]. This is accompanied by increases in the muon spin relaxation rate and the amplitude of the nonrelaxing part of the signal (**Fig.25b**). The magnetic field experienced by muons is estimated to be 0.7 mT at low temperature, around two orders of magnitude smaller than that observed in other pyrochlore iridates [70]. These results suggest that the low-temperature state involves exceptionally small static magnetic moments, $0.01\mu_B$/Ir. The relaxation rate increases further below 0.23(4) K, consistent with an upturn in the specific heat, suggesting the existence of a second low-temperature magnetic transition. Indeed, the coefficients ($\gamma$ and $\beta$) of the low-temperature $T$ and $T^3$ terms of the specific heat C(T) are strongly field-dependent. The state also has a conspicuously large Wilson ratio $R_W \approx 53.5$ and an unusual Hall resistivity that abruptly changes below 80 K without any correlation with the magnetic behavior (**Fig. 26b**) [69]. These unconventional properties, along with the novel behavior observed in metallic hexagonal $SrIrO_3$ [58], define an exotic class of SOI metals (**Fig. 26a**) in which strongly competing interactions induce non-Fermi liquid states that generate magnetic instabilities.

# IV. Double-Perovskite Iridates with $Ir^{5+}(5d^4)$ Ions: Nonmagnetic Singlet $J_{eff} = 0$ State?

The strong SOI limit is expected to lead to a nonmagnetic singlet ground state, which can be simply understood as a $J_{eff} = 0$ state arising from four electrons filling the lower $J_{eff} = 3/2$ quadruplet in materials with $d^4$ ions, such as $Ru^{4+}(4d^4)$, $Rh^{5+}(4d^4)$, $Re^{3+}(5d^4)$, $Os^{4+}(5d^4)$, as well as $Ir^{5+}(5d^4)$ (see the schematic in **Fig. 1d,** and **Fig. 27c)** [17,73,74,75,80,166,167,168,169,170,171,194]. Indeed, the $J_{eff} = 0$ state has been used to explain the absence of magnetic ordering in the pentavalent post perovskite $NaIrO_3$ **[80]**, although it has also been attributed to structural distortions [169]. On the other hand, a low-spin S = 1 state is expected large crystalline fields when U and the Hund's rule coupling $J_H$ are much greater than the SOI $\lambda$, which is a condition commonly seen in ruthenates (**Fig. 27a**) [19]. In this case only the $t_{2g}$-orbitals are relevant, which are all singly occupied except for the one with lowest energy, yielding an effective S=1.

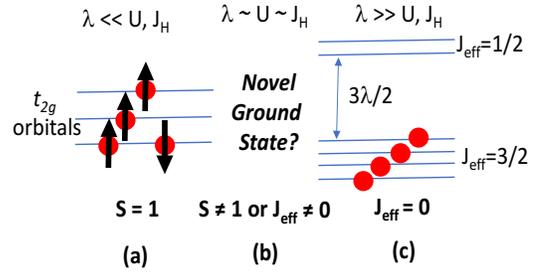

**Fig. 27.** The ground state of $4d^4$ or $5d^4$ ions: **(a)** the low spin state S=1, and **(c)** the singlet $J_{eff}=0$ state and **(b)** an intermediate state between S=1 and $J_{eff}=0$ states.

Nevertheless, theoretical and experimental studies suggest that novel states in these materials can emerge when exchange interactions (0.05-0.10 eV), $J_H$, singlet-triplet splitting (0.050-0.20 eV) and the SOI are comparable, and therefore compete. Exotic states are expected in $d^4$-Mott insulators that support "intermediate-strength" SOI and U [167,168,169,170,195,196].

Pentavalent iridates attracted attention when experimental and theoretical studies showed evidence that contraindicated the anticipated $J_{eff} = 0$ state [75,166,167,168,169]. One surprising experiment addressed a distorted double-perovskite $Sr_2YIrO_6$ that exhibits an exotic magnetic state below 1.3 K rather than an expected $J_{eff} = 0$ state or a S=1 state (**Fig. 27b**) [73]. $Sr_2YIrO_6$ adopts a monoclinic structure essentially derived from the $SrIrO_3$ perovskite by replacing every other Ir by nonmagnetic Y; the remaining magnetic $Ir^{5+}$ ions form a network of edge-sharing tetrahedra or a face centered cubic (FCC) structure with lattice parameters elongated compared to the parent cubic structure, as shown in **Fig. 28**. Because of the differences in valence state and ionic radius between $Y^{3+}$ and $Ir^{5+}$ ions, no significant intersite disorder is expected. This and other related double perovskite iridates have two strongly unfavorable conditions for magnetic order, namely, pentavalent $Ir^{5+}(5d^4)$ ions which are anticipated to have $J_{eff} = 0$ singlet ground states in the strong SOI limit, and geometric frustration in a FCC structure formed by the $Ir^{5+}$ ions (**Fig. 28**).

The emergence of the unexpected magnetic ground state was initially attributed to effects of non-cubic crystal fields on the $J_{eff} = 1/2$ and $J_{eff} = 3/2$ states because such effects were not included in the original model [7,8]. However, there are other possible origins for a magnetic moment in pentavalent Ir systems: As the hopping of the $t_{2g}$ electrons increases, the width of the bands increases and the $J_{eff} = 1/2$ and $J_{eff} = 3/2$ bands may overlap so that the $J_{eff} = 1/2$ state is partially filled and the $J_{eff} = 3/2$ state has a corresponding number of holes. This may result in a magnetic moment. Interactions, in particular, the Hund's rule exchange, $J_H$, and U couple the different orbitals. Recent theoretical studies [167,168,169, 195,196] predict a quantum phase transition with increasing hopping of the electrons from the expected $J_{eff} = 0$ state to a novel magnetic state with local $5d^4$ moments. Furthermore, a band structure study of a series of double-perovskite iridates with $Ir^{5+}(5d^4)$ ions shows that the $e_g$ orbitals play no role in determining the ground state. It confirms the observed magnetic state in distorted $Sr_2YIrO_6$ [169] and also predicts a breakdown of the $J_{eff} = 0$ state in undistorted, cubic $Ba_2YIrO_6$ as well because the magnetic order is a result of band structure, rather than of non-cubic crystal fields in these double-perovskites.

Indeed, a low-temperature magnetic ground state is recently observed in the entire series of $(Ba_{1-x}Sr_x)_2YIrO_6$ [197]. This observed magnetic state, which is similar to that in $Sr_2YIrO_6$, evolves gradually

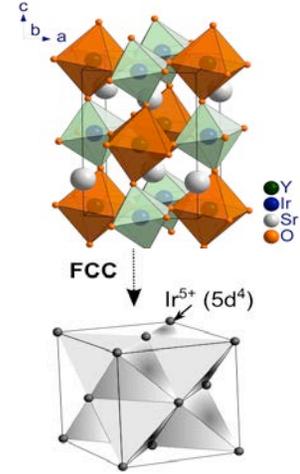

**Fig. 28.** *Upper Panel: The double perovskite crystal structure of $Sr_2YIrO_6$; Lower Panel: The ordered replacement of nonmagnetic Y ions for magnetic Ir ions leading to a FCC lattice with geometrically frustrated edge-sharing tetrahedra formed by the pentavalent $Ir^{5+}$ ions (Ref.73).*

with changes of the lattice parameters while retaining the underlying AMF characteristics (see **Fig.29**). However, the small value of the ordered magnetic moment and small magnetic entropy removal associated with heat capacity anomalies imply that this magnetic state is weak and barely stable compared to either the S = 1 state that is commonly observed among heavy $d^4$ transition element ions, or the $J_{eff}$

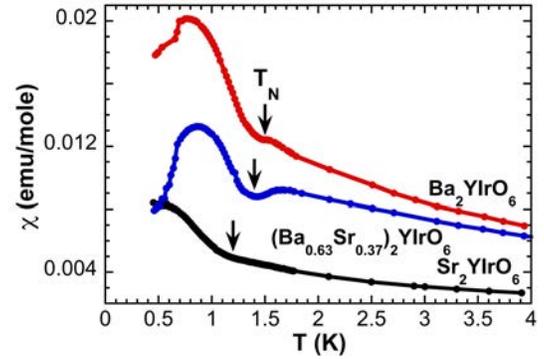

**Fig.29.** *$(Ba_{1-x}Sr_x)_2YIrO_6$: The temperature dependence of magnetic susceptibility $\chi(T)$ at low temperatures (Ref.197).*

= 0 state driven by strong SOI [197]. These circumstances make this magnetic state unique and intriguing. Indeed, a recent resonant inelastic x-ray scattering (RIXS) reveals that the Hund's rule coupling $J_H$ = 0.25 eV for $Ir^{5+}$ in the double perovskite iridates, suggesting that $J_H$ should be treated on equal footing with the SOI in these materials [198]. A more commonly accepted argument is that the SOI competes with a comparable exchange interaction $J_H$ and a generically large electron hopping (because of the extended nature of 5d-orbitals) that suppresses the singlet $J_{eff}$ = 0 ground state. At the same time, the SOI breaks down the low-spin S = 1 state. This "simultaneous" destabilization of the S = 1 and $J_{eff}$ = 0 states leads to a magnetic state that resides somewhere between the other two. The above picture is not without controversy. There are also theoretical and experimental studies that suggest a non-magnetic state in the double-perovskite iridates [74,75,194,199].

While the $J_{eff}$ =1/2 insulating state model successfully captures the new physics observed in many iridates, recent studies suggest that it may not be adequate to describe new phenomena when the relative strength of the SOI critically competes with the strength of electron hopping and exchange interactions. It is worth mentioning once again that the $J_{eff}$ = 1/2 model is a single-particle approach that assumes that SOI is greater than Hund's rule interactions among the electrons. Its validity needs to be closely examined when the Hund's rule interactions among the electrons are comparable to SOI in $d^4$-electron systems [198].

Nevertheless, the anticipated magnetic ground state [167,168,169,195,196] may indicate that the SOI may not be as dominating as initially anticipated, thus leading to magnetic order rather than a singlet ground state in the double perovskite iridates. This magnetic state could be extraordinarily fragile as evidenced in the varied magnetic behavior reported in both experimental and theoretical studies [74,75,75,194,197]. It is clear that the stability limits of the spin-orbit-coupled $J_{eff}$ states in heavy transition metal materials must be investigated.

## V. Future Challenges and Outlook

The SOI is a strong competitor with U and other interactions, and creates an entirely new hierarchy of energy scales in iridates (**Table 2**), which provides fertile ground for the discovery of novel physics. Indeed, a burgeoning list of theoretical proposals and predictions of exotic states, such as the spin liquid state, superconductivity, Weyl semimetals, correlated topological insulators, etc. is truly remarkable and stimulating. It is perhaps even more intriguing that these predictions have met only limited experimental confirmation thus far. The physics of iridates clearly presents urgent intellectual challenges both theoretically and experimentally, and this field is still in its infancy.

Iridates tend to be magnetic insulators, chiefly because of the combined effect of SOI and U; and they also display a number of interesting empirical trends that defy conventional wisdom. In particular, iridates rarely metalize at high pressures, regardless of their crystal structures; however, when they do, the induced metallic state is extraordinary, such as that in $Sr_3Ir_2O_7$ near 60 GPa. On the other hand, a metallic state can be readily induced via slight chemical doping in layered perovskite and hexagonal iridates, but remarkably not in honeycomb iridates. The charge gap and the magnetic state do not necessarily track each other. There are seldom first-order phase transitions in iridates, if any at all: all known transitions induced by temperature, magnetic field, pressure, or chemical doping are gradual and continuous (the sharp transitions in the "S"-shaped IV curves may be a rare exception). Moreover, non-linear Ohmic behavior and/or switching effects are also commonplace in iridates. While all these curious properties pose tantalizing prospects for unique functional materials and devices, they also pose a series of intriguing questions that may provide the impetus for advancing our understanding of this class of materials:

- *Is the $J_{eff}=1/2$ insulating state ubiquitous in other 5d-materials (even in non-cubic crystal fields and/or with non-negligible Hund's rule interactions between the electrons)?*

- *Why do iridates seldom metallize at high pressures?*

- *In contrast, why can even small chemical doping readily induce a metallic state in most insulating iridates (except for honeycomb iridates)?*

- *Why does superconductivity remain elusive in $Sr_2IrO_4$ despite its similarities to the cuprates, and extensive theoretical and experimental efforts addressing superconductivity in this model material?*

- *Will driving the hidden non-dipolar magnetic order to its quantum critical point provide a path toward generating the elusive superconducting state in $Sr_2IrO_4$?*

- *Why do all honeycomb iridates and ruthenates magnetically order in a similar temperature range, despite their different SOI?*

- *Can we go beyond discussions of Mott, Mott-Hubbard and Slater insulators to better understand the role of the magnetic transition in the formation of an insulating state in iridates?*

- *A metallic state is rare in iridates, but when it does happen it has extraordinary properties. How do we describe it?*

- *How do we explain current-sensitive transport properties or non-linear Ohmic behavior, in particular, the S-shaped I-V characteristic?*

- *Does the nonmagnetic singlet $J_{eff} = 0$ state exist in iridates with $Ir^{5+}(5d^4)$ ions?*

- *More generally, how do we describe the Mott insulating state with "intermediate-strength" or strong SOI in materials with $d^4$ ions, such as $Ru^{4+}(4d^4)$, $Rh^{5+}(4d^4)$, $Re^{3+}(5d^4)$, $Os^{4+}(5d^4)$ and $Ir^{5+}(5d^4)$?*

It needs to be reiterated that one defining characteristic of iridates is that the strong SOI renders a strong coupling of physical properties to the lattice degrees of freedom, which is seldom seen in other materials. The wide range of novel phenomena and the limited experimental confirmation for many predicted ground states imply a critical role of subtle, local lattice properties that control ground states, and needs to be more adequately addressed both experimentally and theoretically.

Because of the unique distortions of the Ir-O-Ir bond in iridates, the lattice properties, and therefore physical properties, can be readily tuned via perturbations such as magnetic field, electric field, pressure, epitaxial strain or chemical doping. That transport properties sensitively change with electrical current or electric field (e.g., the S-shaped IV curves) is particularly promising in terms of fundamental research and development of potential functional devices, since the application of an electric field is much more convenient than the application of a magnetic field or pressure.

Of course, the nature of the extraordinary structural sensitivity of iridates also calls for extraordinarily high-quality single crystals and epitaxial thin-films. Moreover, the strong neutron absorption caused by the high atomic number of iridium, 77, requires not only high-quality but also large and preferably plate-like, bulk single crystals for more definitive magnetic studies such as inelastic neutron scattering. This is particularly urgent for a number of iridates such as pyrochlore iridates, $Li_2IrO_3$, $Sr_3Ir_2O_7$ and double perovskites with $Ir^{5+}$ ions, whose magnetic structures are yet to be conclusively determined. This is a daunting experimental challenge because iridates tend to have both high melting points and high vapor pressure, which require sophisticated techniques for single-crystal synthesis.

Finally, we would like to reiterate that it is impossible to review such a broad and rapidly evolving field in a single article. We can only provide a basic introduction, some general remarks and observations, and a number of hopefully instructive examples.

**Acknowledgments** This work is supported by the US National Science Foundation via grants DMR-1265162 and DMR-1712101. PS is supported by the US Department of Energy (Basic Energy Sciences) under grant No. DE-FG02-98ER45707. GC is deeply indebted to Drs. Lance De Long, Feng Ye, Daniel Haskel, Hae Young Kee and David Hsieh for their insightful comments and suggestions, which significantly improve this *Review*.